\documentclass[iop]{emulateapj}

\usepackage{amsmath}

\usepackage{graphicx}
\shorttitle{Evolution of Normal Galaxy XLF}
\shortauthors{M. Tremmel et al.}

\begin{document}

\title{Modeling the Redshift Evolution of the Normal Galaxy X-ray Luminosity Function}

\author{ M.\ Tremmel$^{1}$, T.\ Fragos$^{2}$, B.D.\ Lehmer$^{3,4}$, P.\ Tzanavaris$^{3,4}$,  K.\ Belczynski$^{5,6}$, V.\ Kalogera$^{7}$, A. R.\ Basu-Zych$^{4}$,  W.\ M.\ Farr$^{7}$, A.\ Hornschemeier$^{4}$,  L.\ Jenkins$^{4}$, A.\ Ptak$^{4}$, A.\ Zezas$^{8,9,2}$} 

\altaffiltext{1}{Department of Astronomy, University of Washington, Box 351580, U.W., Seattle, WA 98195-1580, USA}
\altaffiltext{2}{Harvard-Smithsonian Center for Astrophysics, 60 Garden Street, Cambridge, MA 02138 USA}
\altaffiltext{3}{The Johns Hopkins University, Homewood Campus, Baltimore, MD 21218, USA}
\altaffiltext{4}{NASA Goddard Space Flight Centre, Code 662, Greenbelt, MD 20771, USA}
\altaffiltext{5}{Astronomical Observatory, University of Warsaw, Al. Ujazdowskie 4, 00-478 Warsaw, Poland}
\altaffiltext{6}{Center for Gravitational Wave Astronomy, University of Texas at Brownsville, Brownsville, TX 78520, USA}
\altaffiltext{7}{Center for Interdisclipinary Research and Exploration in Astrophysics and Department of Physics and Astronomy, Northwestern University, 2145 Sheridan Road, Evanston, IL 60208, USA}
\altaffiltext{8}{Department of Physics, University of Crete, P.O. Box 2208, 71003 Heraklion, Crete, Greece}
\altaffiltext{9}{IESL, Foundation for Research and Technology, 71110 Heraklion, Crete, Greece}
\email{mjt29@astro.washington.edu}

\begin{abstract}
	
Emission from X-ray binaries (XRBs) is a major component of the total X-ray luminosity of normal galaxies, so X-ray studies of high redshift galaxies allow us to probe the formation and evolution of X-ray binaries on very long timescales ($\sim 10$ Gyr). In this paper, we present results from large-scale population synthesis models of binary populations in galaxies from $z = 0$ to $\sim 20$. We use as input into our modeling the Millennium II Cosmological Simulation and the updated semi-analytic galaxy catalog by Guo et al. (2011) to self-consistently account for the star formation history (SFH) and metallicity evolution of each galaxy. We run a grid of 192 models, varying all the parameters known from previous studies to affect the evolution of XRBs. We use our models and observationally derived prescriptions for hot gas emission to create theoretical galaxy X-ray luminosity functions (XLFs) for several redshift bins.  Models with low CE efficiencies, a 50\% twins mass ratio distribution, a steeper IMF exponent, and high stellar wind mass loss rates best match observational results from Tzanavaris \& Georgantopoulos (2008), though they significantly underproduce bright early-type and very bright ($L_x > 10^{41}$) late-type galaxies.  These discrepancies are likely caused by uncertainties in hot gas emission and SFHs, AGN contamination, and a lack of dynamically formed Low-mass XRBs.  In our highest likelihood models, we find that hot gas emission dominates the emission for most bright galaxies.  We also find that the evolution of the normal galaxy X-ray luminosity density out to z = 4 is driven largely by XRBs in galaxies with X-ray luminosities between $10^{40}$ and $10^{41} erg ~s^{-1}$.

\end{abstract}


\maketitle

\section{Introduction}

X-ray binaries (XRBs) are believed to be major contributors to the overall X-ray luminosity of normal galaxies \citep[those not dominated by the emission of a nuclear supermassive black hole;][]{1989ARA&A..27...87F,2003ApJ...586..826K,2006ARA&A..44..323F}.  Normal early-type galaxies have older stellar populations and their X-ray emission is dominated by low-mass XRBs (LMXBs) and hot interstellar medium (ISM).  On the other hand, the X-ray emission of normal late-type galaxies, which are still actively forming stars, have significant contributions from both LMXBs and high mass XRBs (HMXBs).
	
	X-ray and multiwavelength studies of galaxies using \textit{Chandra} and \textit{XMM-Newton} have yielded a great deal of information about the X-ray luminosities of galaxies, including many X-ray correlations that have been established to hold out to at least $z=1$ \citep[e.g.][]{2007ApJ...657..681L,2008ApJ...681.1163L,2011MNRAS.417.2239S,2012MNRAS.420.2190V,2012arXiv1210.3357B,2012ApJ...748...50C}.  These relations include a strong correlation between X-ray  emission from HMXBs and the star formation rate (SFR) of galaxies \citep[e.g.][]{2003A&A...399...39R,2004MNRAS.347L..57G,2010ApJ...724..559L,2012MNRAS.419.2095M} as well as a scaling relation between the emission from LMXBs and the stellar mass of a galaxy \citep{2004MNRAS.347L..57G,2004MNRAS.349..146G,2010ApJ...724..559L,2011ApJ...729...12B,2012arXiv1202.2331Z}.  Recent ultradeep \textit{Chandra} and multiwavelength surveys \citep[e.g][]{2005ARA&A..43..827B} have allowed for robust tests of these relations in very distant galaxies.  For example, \citet{2012arXiv1210.3357B} use a 4 Ms exposure of CDF South \citep{2011ApJS..195...10X} and X-ray stacking to study faint X-ray sources out to $z\sim8$, finding that the relation between X-ray production and star formation rate undergoes a small amount of evolution out to $z\sim4$ that is likely driven by the metallicity evolution of HMXBs.
	
	Galaxy XLFs derived from recent observations show significant evolution with redshift \citep{2004ApJ...607..721N, 2005A&A...440...23R,2007ApJ...667..826P,2008A&A...480..663T}.  \citet{2008A&A...480..663T} (T\&G08 hereafter) use data from three \textit{Chandra} deep fields (CDFS, E-CDFS, and CDFN) and the wide area survey XBootes to compile observations of 207 X-ray luminous normal galaxies (101 early-type and 106 late-type)  out to $z \sim 1.4$.  They find a clear evolution of the galaxy XLF normalization with redshift that is driven almost exclusively by late-type galaxies.  More specifically, this evolution is proportional to $(1+z)^{k}$ with $k=2.2 \pm 0.3$ for the total population, $k=2.4^{+1.0}_{-2.0}$ for late-type galaxies, and, for the early-type population, $k=0.7^{+1.4}_{-1.6} $ (consistent with zero).  Because XRBs are major contributors to the total X-ray emission of normal galaxies, observationally derived XLFs can put constraints on theoretical models of XRB formation and evolution. 
	
	At present, there has been little theoretical work done on the evolution of XRB populations over cosmological timescales \citep{1998ApJ...504L..31W, 2001ApJ...559L..97G, 2011ApJ...733....5Z}.  It is thought that XRBs could play a major role in the evolution of these XLFs \citep{1998ApJ...504L..31W, 2001ApJ...559L..97G, 2004ApJ...607..721N}.  \citet{2001ApJ...559L..97G}, using a semi-empirical, semi-analytical approach, linked XRB lifetimes with star formation rates (SFRs), showing that SFRs that are evolving on cosmological timescales significantly affect the XRB populations and, therefore, the integrated galactic X-ray emission.  This predicted evolution should be evident even at lower redshifts \citep[$z \lesssim 1 $;][]{1998ApJ...504L..31W}.
	
	Recently, the advances in available multi-wavelength observations of distant galaxies, as well as our understanding of binary stellar evolution and galaxy formation and evolution, have reached a level of maturity that allows us to conduct an in-depth study of the XRB populations of distant galaxies.  In this paper, we study the evolution of X-ray binaries on cosmologically significant timescales, using data from detailed, large scale simulations.  We use data from a catalog created by \citet{2011MNRAS.tmp..164G} using semi-analytical galaxy evolution models applied to the recent \textit{Millennium Cosmological Simulation}.  These data are used in tandem with the binary population synthesis (PS) code, \textit{StarTrack}, to simulate the XRB populations of individual galaxies from $z=0$ to $\sim20$, taking into account the full star formation and merger histories of each galaxy.  From these models, we derive the integrated X-ray emission of each galaxy and compare the resulting galaxy XLFs and their evolution to observed galaxy XLFs.  Our goal is to obtain better constraints on the parameter space for models of XRB formation and evolution, and to better understand the nature of the X-ray emission of galaxies at high redshifts.
	
	Recently, \citet{Fragos2012} (F12 hereafter) used similar techniques as those described in this paper to study the evolution of the global XRB population with redshift.  They model how the total Universal specific X-ray luminosities ($L_{\rm X}$ per unit stellar mass, star formation rate, and volume) of LMXBs and HMXBs evolve over cosmic time out to $z \sim 20$.  Their models were constrained by observed luminosities of HMXB and LMXB populations in the local Universe.  They found that the LMXB population dominates the total population at low redshifts, with HMXB contributions becoming dominant for redshifts higher than $z \sim 3.1$.
	
	The outline of the paper is as follows.  Section 2 describes our simulation tools, \textit{StarTrack} and \textit{The Millennium Simulation II}, and the methodology we follow in developing our models of XRB populations in galaxies.  Section 3 describes how we compare our models to observational results, namely those of T\&G08, and the statistical analysis we use to determine our best models.  In section 4, we describe and discuss our results, and we conclude with a summary in section 5.
	
	\section{Simulating X-ray Luminosities of Galaxies}
\subsection{The Millennium Cosmological Simulation}

	The \textit{Millennium Cosmological Simulation} is an unprecedented computational effort to simulate the dark matter distribution in the Universe  \citep[see][for details]{2005Natur.435..629S}.  In this study, we use the data from the most recent Millennium Run II  \citep[MRII hereafter;][]{2009MNRAS.398.1150B}. This is an N-body simulation that follows the evolution of $2160^{3}$ particles each of mass $6.9 \times 10^{6} h^{-1}M_{sun}$ within a co-moving box with sides each of size $100 h^{-1} Mpc$.  The cosmological model used in the simulation is a $\Lambda$CDM model with $\Omega_{m} = 0.25$, $\Omega_{\Lambda} = 0.75$, $\Omega_{b} = 0.045$, and $h = H_{0}/100$ km/s/Mpc = 0.73. 
	
	The MRII has 60 snapshots in time that were saved and analysis was done to identify substructures within the dark matter distributions, including dark matter halos.  \citet[][G11 hereafter]{2011MNRAS.tmp..164G} use a semi-analytic procedure to track the evolution of the galaxies that exist within these halos.  Once subhalos are identified, their merger trees are derived.  The evolution of these subhalos provide the base for the galaxy formation model.  The models used by G11 build upon the work of \citet{2007MNRAS.375....2D}, making improvements to the treatment of supernova feedback, reincorporation of ejected gas, galaxy sizes, the distinction between satellite and central galaxies, and the effect of the environment on galaxies.  While semi-analytical models do not supply accurate details about individual galaxies, they are very useful for understanding general characteristics of large populations of galaxies.  These semi-analytical models, when applied to the MRII simulation, are able to accurately reproduce observed characteristics of galaxy populations, e.g.,  the abundance and large-scale clustering of low-z galaxies, the Tully-Fisher relation, stellar mass and luminosity functions of low-z galaxies, the halo-galaxy mass relationship, and the evolution of the cosmic star formation density.  However, these models overproduce passive low mass galaxies and fail to reproduce the observed abundances, clustering, and mass functions of high redshift ($z > 1.0$) galaxies. In this paper, we will be comparing with X-ray observations of galaxies out to redshift 1.4, which is in a regime where the G11 model is still fairly accurate.
		
	The result of G11's semi-analytic model is a catalog of the galaxy population at 60 different times between $z \sim 20$ (about 13.4 billion years ago) and the present day.  These catalogs include properties such as metallicity, stellar mass, bulge mass, the mass of hot and cold gas, rest frame luminosity magnitudes, etc., for each of the galaxies in the simulation box, as a function of time.

\subsection{StarTrack}
To simulate the XRB populations of the galaxies from MRII, we use \textit{StarTrack}, a current binary population synthesis code that has been tested and calibrated using detailed binary star calculations and incorporates all the most important physical processes of binary evolution \citep{2002ApJ...572..407B,2008ApJS..174..223B}: \\
	(i) The evolution of single stars and non-interacting binary components, from their birth, taken as the time of their initial emergence onto the main sequence, to compact remnant formation using evolutionary formulae of \citet{2000MNRAS.315..543H} modified as described in \citet{2008ApJS..174..223B}.   Various wind mass loss rates and their effect on stellar evolution are also incorporated into the code and have been recently updated \citep{2010ApJ...714.1217B}\\
	(ii) The time evolution of orbital properties.  \textit{StarTrack} numerically integrates a set of four differential equations describing the evolution of orbital separation, eccentricity, and component spins, taking into account tidal interactions, magnetic braking, gravitational radiation and stellar wind mass losses. \\
	(iii) All types of mass-transfer phases.  This includes both stable and unstable mass-transfer processes, which are driven by either nuclear evolution or angular momentum loss.  Unstable mass transfer is encountered most often as a direct consequence of rapid stellar expansion during nuclear evolution, but angular momentum loss may also lead to instability. \\
	(iv) Supernova explosions, which are treated by taking into account mass/angular momentum losses as well as supernova asymmetries (through natal kicks to neutron stars and black holes). \\
	(v) X-ray emission, which is tracked for accreting binaries with compact object primaries (both for wind-fed and Roche-lobe overflowing systems).  The resulting X-ray luminosities are calculated from the secular averaged mass accretion rate, but are not calculated for unstable accretion phases because the timescales are very short.
	
	The models in this paper include a recent revision of \textit{StarTrack} that incorporates updated stellar winds and their re-calibrated dependence on metallicity \citep{2010ApJ...714.1217B}.  However, two more recent upgrades have not been incorporated into these results, as the simulations were run long before the changes were made.  The first update includes a revised neutron star and BH mass distribution based on fully consistent supernova simulations \citep{2011arXiv1110.1635B,2012ApJ...749...91F}.  The second, most recent upgrade improves upon the treatment of donor stars in CE events via usage of the actual value of the $\lambda$ parameter, the measure of the central concentration of the donor and envelope binding energy, for which usually a constant value is assumed \citep{2012arXiv1202.4901D}.
	
	Table 1 lists the input parameters  of our population synthesis models, which can be put into two categories.  In the first category there are the parameters that correspond to the initial properties of the binary population.  These values are relatively well constrained by the most recent observational surveys.  Also in the group are stellar wind prescriptions and natal kick distributions, which can also be constrained by observations.  In the second group are the truly ``free'' parameters that correspond to poorly understood physical processes, which we are not able to model in detail.  One of these truly ``free'' parameters is the common envelope (CE) efficiency ($\alpha_{CE}$), which measures how efficiently orbital energy loss is  transformed into thermal energy that will expel the donor's envelope during the CE phase.We note that in our calculations we combine $\alpha_{\rm CE}$ and $\lambda$, the binding energy parameter described above, into one CE parameter. Whenever we mention the CE efficiency $\alpha_{\rm CE}$, we refer in practice to the product $\alpha_{\rm CE} \times\lambda$, effectively treating $\alpha_{\rm CE} \times \lambda$ as a free parameter \citep[see][for details]{2008ApJS..174..223B}.  
	
	We create a grid of 192 PS models, a subset of those used in F12, run for nine different metallicities and each simulating $5.12 \times 10^{6}$  stars for 14 Gyr. In this grid, we varied all the parameters known from previous studies to affect the evolution of XRBs and the formation of compact objects in general \citep{2007ApJ...662..504B,2010ApJ...714.1217B,2008ApJ...683..346F,2010ApJ...719L..79F,2009ApJ...699.1573L}.   Specifically, we vary the CE efficiency, initial binary mass ratio distribution, initial mass function (IMF), supernova (SN) kicks for direct collapse BHs, and stellar wind strength.  We also take into account the possibility of CE inspirals with Hertzsprung gap donors that could terminate binary evolution barring the subsequent XRB formation \citep{2007ApJ...662..504B}.
		
	For all models, we assume a Maxwellian distribution of supernova kicks given by \citet{2005MNRAS.360..974H}, with $\sigma = 265 km/s $.  For compact objects formed with partial mass fallback, the natal kicks given by the \citet{2005MNRAS.360..974H} distribution are decreased by a factor of $(1-f_{fb})$, where $f_{fb}$ is the fraction of the stellar envelope that falls back after the SN explosion.  In our standard prescription, direct collapse (DC) BHs, BHs formed with $f_{fb} = 1$, are given no natal kick.  However, due to recent theoretical evidence that even the most massive stellar black holes have probably received small asymmetric kicks \citep{2010ApJ...725.1984L,2010Natur.468...77V}, in some models used in this work we set a lower limit (0.1) on the amount by which the natal kicks may be decreased due to mass fallback, allowing for small natal kicks to be given to direct collapse BHs.  
	
	The mass of the primary star in each binary is determined by the adopted IMF.  It is important to note here that, because we sample the IMF with only the primary star, we are only sampling the high mass end of the IMF because the primary stars that form XRBs must be massive enough to form a BH or NS.  The mass of the secondary star is calculated using a distribution function for the binary mass ratio, $q = M_{secondary}/M_{primary}$.  We vary the distribution of $q$ between a flat distribution in the range $q=0$ to $1$ and a distribution that has 50\% of the binaries follow a distribution with  $q=0$ to $1$ and the other half follow a  ``twins'' distribution, with $q=0.9$ to $1$.
	
	The models and their numbers used in this work are the same as those used in F12, except here we exclude the models 97-192, which have all systems following the pure ``twins'' q-distribution.   \citet{Fragos2012}	show that models with the pure ``twins'' distribution are unphysical, as they fail to reproduce local populations of XRBs.  Thus, we exclude them here.
		
	We want to note that our population synthesis code calculates the bolometric luminosity of each XRB based on the rate of mass transfer. In order to compare our model results with observed datasets we need to estimate the X-ray luminosity in a specific energy band, which in this study is the soft X-ray band of \emph{Chandra} ($0.5-2.0\, keV$). In order to calculate the bolometric correction, we used two sets of published X-ray spectra from Galactic neutron star and black hole XRBs at different spectral states \citep{2006ARA&A..44...49R, 2010ApJ...718..620W}.   Following the same procedure outlined in F12, we derive to the bolometric correction factors for different types of XRBs and use these results to estimate the  $0.5-2.0\, keV$ X-ray luminosity of our modeled XRB population.
	
	We also note that the PS models used here take into account only binary systems formed in the field and not those formed via dynamical interactions in dense stellar systems.  Dynamical formation in globular clusters (GC), for example, is a significant formation pathway for LMXBs in old, GC rich elliptical galaxies \citep[e.g.][]{2008ApJ...689..983H,2012arXiv1202.2331Z}.  Further, the LMXB populations formed in GCs can be as much as $2-3$ times more luminous than the field population in bright elliptical galaxies \citep{Irwin2005}.
	
	For each model and each metallicity value, we calculate the X-ray luminosity as a function of age for single bursts of star-formation, where we also take into account the effect of transient XRBs \citep[see ][]{2008ApJ...683..346F, 2009ApJ...702L.143F}.  Taking into account the assumed initial mass function and initial mass ratio distribution, we normalize the total X-ray luminosity to a nominal population of $10^{10}\, M_{\odot}$. This quantity, $L_{X,spec}(t)  (erg/s/10^{10}M_{\odot})$, is the \textit{specific X-ray luminosity} of a single age stellar population as a function of its age. The specific X-ray luminosity coming from our population synthesis models can be convolved with the star-formation history and metallicity evolution of a galaxy to calculate the total X-ray luminosity of its complex stellar population.
			
\begin{deluxetable*}{lccc} 
\centering
\tablecolumns{4}
\tabletypesize{\scriptsize}
\tablewidth{0pt}
\tablecaption{Model Parameters
\label{modelparam}} 
\tablehead{ \colhead{Parameter} & 
     \colhead{Notation} & 
     \colhead{Value} &
     \colhead{Reference} 
     }
\startdata
Initial Orbital Period distribution\tablenotemark{$\dagger$} & F(P) 				& flat in logP 													& \citet{1983Abt}\\
Initial Eccentricity Distribution\tablenotemark{$\dagger$} 	& F(e)		 		& Thermal $F(e)\sim e$ 											& \citet{1975MNRAS.173..729H}\\
Binary Fraction\tablenotemark{$\dagger$}					& $f_{\rm bin}$ 	& 50\% 															& \\
Magnetic Braking\tablenotemark{$\dagger$} 					&					& 																& \citet{2003ApJ...599..516I} \\
Metallicity\tablenotemark{$\dagger$}           				& $Z$           	& 0.0001, 0.0002, 0.005, 0.001,  								& \\
									&					& 0.002, 0.005, 0.01, 0.02, 0.03								& \\
IMF  (slope)\tablenotemark{$\dagger$} 						&      				&  -2.35 or  -2.7 												&\citet{2001MNRAS.322..231K,2003ApJ...598.1076K}\\
Initial Mass Ratio Distribution\tablenotemark{$\dagger$} 	& F(q) 				& Flat, twin, or 50\% flat plus 50\% twin 						&\citet{2007ApJ...670..747K,2006ApJ...639L..67P}\\
CE Efficiency\tablenotemark{$\star$} 						& $\alpha_{\rm CE}$ &  0.1, 0.2, 0.3, or 0.5 										& \citet{2003MNRAS.341..385P}\\
Stellar wind strength\tablenotemark{$\dagger$} 				& $\eta_{\rm wind}$ &  0.25, 1.0, or 2.0 											&  \citet{2010ApJ...714.1217B}\\
CE during HG\tablenotemark{$\star$} 						& 					& Yes or No 													& \citet{2007ApJ...662..504B}\\
SN kick for ECS/AIC\tablenotemark{a} NS\tablenotemark{$\dagger$}				& 					& 20\% of normal NS kicks	 									& \citet{2009ApJ...699.1573L}\\
SN kick for direct collapse BH\tablenotemark{$\star$} 		& 					& Yes or No 													& \citet{2010ApJ...719L..79F}

\enddata
\label{model_param}
\tablenotetext{$\dagger$}{Observationally constrained parameters}
\tablenotetext{$\star$}{``Free'' parameters}
\tablenotetext{a}{Electron Capture Supernova / Accretion Induced Collapse}
\end{deluxetable*}

	
	\subsection{Convolving \textit{StarTrack} with the G11 catalog}
	
	The MRII catalog created by G11 corresponds to 60 snapshots that span a redshift range from $z=0$ to $\sim 20$.  For each galaxy, we can derive its complete progenitor tree.  Each progenitor galaxy has a unique SFR and metallicity, so for every stellar population in a target galaxy we know during what timeframe and at what metallicity that population was created.  We then convolve the star formation histories with $L_{X,spec}(t)$ derived from the PS models for the appropriate metallicity values.  
	
	The SFRs given for each galaxy in the G11 catalog are averaged over the entire timestep, $\Delta t $, between subsequent snapshots so that the total new stellar mass created in a given progenitor galaxy is $M_{new} = SFR_{prog} \Delta t$. In order to account for the possibility of starbursts, we assume that all new stellar mass forms in a 20 Myr burst occurring at a random time between subsequent snapshots, $t_{i} < t_{burst} < t_{i+1}$, where $t_i$ is the timestamp associated with the snapshot of a given progenitor galaxy.  This effect is important only for the HMXBs of young populations, and the effect on LMXBs is minimal since their evolution occurs on timescales much longer than the timesteps between snapshots.  A 20 Myr duration is reasonable, given that the most cited values for starburst durations are around 3-10 Myr \citep[e.g.][]{2000ApJ...539..641T,2004ApJ...603..503H}, though there is evidence for longer burst durations on the order of a few $10^{8}$ yrs \citep[e.g.][]{2010ApJ...724...49M}.
	
	By summing the soft band X-ray luminosities of all the stellar populations in a given galaxy, we derive the integrated X-ray luminosity from XRBs.  The end result is a catalog of integrated XRB luminosities of galaxies within the MRII co-moving volume from $z=0$ to $z\sim20$.
	
	\section{Comparing with Observations}
	
	\subsection{Galaxy Classification}
	
	When comparing our results to the observations of T\&G08, we want to distinguish between early and late-type galaxies.  For their sample, T\&G08 cross correlate with other surveys to obtain optical counterparts for their X-ray selected galaxies, which they used for classification.  
	
	The classification of a galaxy as early or late-type can be based either on its morphology or its spectroscopic properties.  In many observational surveys of distant galaxies \citep[e.g. GOODS;][]{2004ApJ...600L..93G}, the morphologies of most galaxies that are observed cannot be determined due to inadequate spatial resolution, and therefore colors are used instead.
	
	For color classification of our model galaxies as early or late-type, we adopt the method developed by \citet{2004ApJ...608..752B}.  They showed that it is possible to define the population of early-type galaxies empirically by using the bimodality of the color distribution, which they studied out to $z\sim1$.  The MRII database includes absolute rest frame magnitudes in the SDSS \textit{ugr} filters, which can easily be transformed to the UBV filters used in \citet{2004ApJ...608..752B}.  The magnitudes include the effects of dust extinction.  Following the \citet{2004ApJ...608..752B} prescription, we define early type galaxies to be galaxies where $\left \langle U-V \right \rangle \ge 1.15 - 0.31z - 0.08*(M_{V} - 5log_{10}h + 20)$.  Figure 1 shows plots of $\left \langle U-V \right \rangle$ vs. $M_{V} - 5log_{10}h$ for the MRII galaxies at various redshifts with the cutoff function overlaid on top and the bimodality is clearly present.
	
	If we instead define galaxy types based on morphology with late type galaxies having $\frac{M_{bulge}}{M_{total}} < 0.7$ and early-type galaxies  $\frac{M_{bulge}}{M_{total}} > 0.7$, we find that there is approximately a $1\%$ and $0.5\%$ contamination among the color defined late and early-types respectively.  Thus, these methods give nearly identical results, but using colors to define morphology allows us to better simulate observations.  The morphology method of classification, since it is independent of color, provides a check on our color classification.  The fact that they both yield similar results is encouraging and indicates that the colors provided by the G11 catalog are able to yield accurate morphology classifications.
	
	\begin{figure*}
	\centering
	\includegraphics{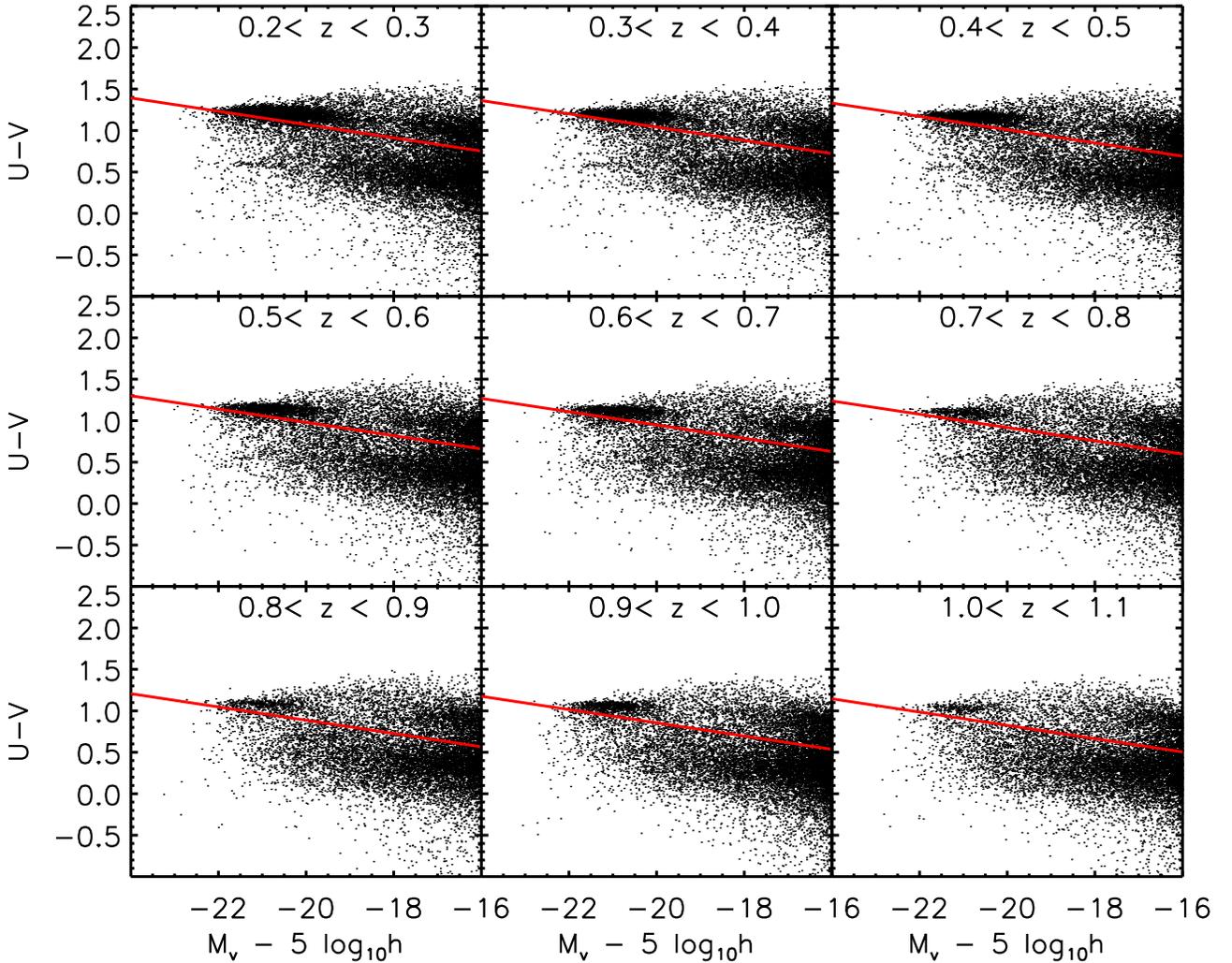}
	\caption{\label{Galaxy Color-Magnitude Diagram} The rest-frame $U - V$ color of simulated galaxies against the absolute magnitude in V-band, $M_{V} - 5log_{10}h$.  It is clear that the G11 galaxy catalog exhibits the same bimodality as observed galaxies.  The red line corresponds to $1.15 - 0.31z - 0.08*(M_{V} - 5log_{10}h + 20)$. This may be compared to the red sequence fitting of U-V colors by \citet{2004ApJ...608..752B}; see their figure 1.  Galaxies that lie above the red line are considered to be early-type galaxies and those that lie below are late-type galaxies.}
	\end{figure*}
	
	\subsection{Creating Model X-ray Luminosity Functions}
		
	For our analysis, we only select galaxies with stellar masses greater than $10^{5}$ solar masses, as galaxies with mass less than this are very unlikely to have X-ray luminosities that are observationally relevant.  For instance, the dwarf galaxies in the SINGS sample, with masses $\sim 10^{7} M_{\odot}$, generally have X-ray luminosities below $10^{37}$ ergs / s (0.5 - 8 keV) with many having no binaries detected at all above $~5\times10^{36}$ ergs / s (0.5 - 8 keV) (Jenkins, et. al. 2012 [in prep]).  The G11 catalog contains several hundred of these more massive galaxies at very high redshift ($z = 19.9$) and over 2 million at $z=0$.  It should also be noted that we assume that all of the galaxies in our sample are normal galaxies.  This is a valid assumption because galaxies with bright AGN only constitute $\sim 2-4 \%$ of all galaxies and, therefore, any selection effects on our data would be minimal \citep{2010ApJ...720..368X, 2010ApJ...723.1447H, 2009ApJ...695..171S}.  Lower luminosity AGN have been found in a much higher percentage ($\sim30-40$\%) of LINER galaxies \citep{1997ApJ...487..568H}.  However, since LINER galaxies themselves make up only $\frac{1}{5}$ to $\frac{1}{3}$ of all galaxies \citep{1997ApJ...487..568H}, this effect is also rather minimal.  Additionally, since this is for lower luminosity AGN, it is likely that our more luminous galaxies would still be dominated by hot gas and XRB emission and would be classified as a normal galaxy.  For example, \citet{2006ApJ...647..140F} find that AGN in their sample of LINER galaxies contribute only 60\% of the 0.5-10 keV luminosity when only considering the central regions of the galaxies.
	
	To compare our results with observations, we derive the XLF for our galaxies by calculating the number density of galaxies versus their integrated X-ray luminosity.  The simulation data are all within a single co-moving volume of constant size.  For the time slice represented in each snapshot, $\phi(L)$ is defined as:
	\begin{equation}
	\phi(L) = \frac{N(L_{min},L_{max})}{V_{MRII}\delta_{L}} 
	\end{equation}
Here, $L_{min}, L_{max}$ are the bin limits, $V_{MRII}$ is the volume of the MRII simulation, which is $(100 \ Mpc/h)^{3}$, and $\delta_{L}$ is the size of the luminosity bin in log space, i.e. $\delta_{L} = Log_{10}(\frac{L_{max}}{L_{min}})$.  Herein lies a major difference between the theoretical and observational luminosity functions.  Observational surveys study a range of redshifts within a light cone.  An entire volume of space cannot be observed at a constant redshift so a range of redshifts is explored. Thus, when calculating $\phi(L)$, observers such as T\&G08 use methods like the one found in \citet{2000MNRAS.311..433P} which uses the following definition:
	\begin{equation}
	\phi(L,z) = \frac{N}{\int_{L_{min}}^{L_{max}} \int_{z_{min}(L)}^{z_{max}(L)} \frac{dV}{dz}\,dz\,dL}
	\end{equation}
Here, $ \frac{dV}{dz}$ represents the rate of change of the survey volume with respect to redshift and $z_{max}(L), z_{min}(L) $ are the redshift ranges for a source as a function of luminosity such that it stays within the flux limits of the survey and within the redshift interval.  Our simulated galaxies, on the other hand, exist within a co-moving volume and our snapshots capture all galaxies that exist at a given redshift.  In order to compensate for this difference, we adopt similar redshift intervals as used by T\&G08 and calculate the XLF for each redshift individually using equation (1) above.  This gives us $\phi(L,z)$.  Then, for each luminosity bin centered at $L$, we take the average value of $\phi(L,z)$ over all $z$ in the interval, giving us an estimate for $\phi(L)$ for that bin in that redshift interval.

		\subsection{X-ray Luminosity from Hot Gas}
	
	In addition to XRBs, the hot ISM in a galaxy can have a significant contribution to its overall X-ray luminosity.  T\&G08 do not distinguish between emission from XRBs and hot gas in their analysis of X-ray bright galaxies.  Their analysis is done in the soft X-ray band, so emission from the hot ISM becomes important and needs to be taken into account when calculating the total integrated luminosities of our galaxies.
	
	We use observationally derived relations to estimate the X-ray luminosity of hot gas in early-type galaxies from their K-band luminosity.  It has been shown that there is a power law relationship between the X-ray luminosity of the hot ISM in early-type galaxies and both the K-band luminosity ($L_{\rm X}\propto L_k^{1.935}$) of the galaxy and the temperature of its hot gas \citep[$T\propto L_{\rm X}^{0.214}$;][]{2011ApJ...729...12B}.  The K-band luminosities of MRII early-type galaxies are easily calculated from mass and age using synthetic stellar population models \citep{2008A&A...484..815B}.  With these relations, we estimate both the full band X-ray luminosity and the temperature of the hot gas in each early-type galaxy.  The spectrum of  hot diffuse gas is assumed to be that of a collisionally ionized diffuse gas as calculated by the APEC XSPEC model and the ATOMDB code \citep{2012ApJ...756..128F}.  The gas temperature estimate from the empirical relations is used as input to the APEC model in order to calculate the luminosity of the hot gas in the soft X-ray band
	
	For late-type galaxies, we estimate the hot gas X-ray luminosity based on the SFRs given in the G11 catalog and the power-law relationship catalog between the soft-band ($0.5-2 keV$) X-ray luminosity of the hot ISM and the SFR for late-type galaxies \citep[$L_{\rm X}\propto SFR^{1.07}$;][]{2004ApJS..151..193S, 2004ApJ...606..829S}.  
	
	In summary, we estimate the total X-ray emission from hot gas in all of the galaxies in the G11 catalog.  We add those values to each galaxy's integrated X-ray emission from XRBs, calculated using \textit{StarTrack}, to obtain the total integrated X-ray luminosity of each galaxy.  We find that on average XRB emission contributes to $\sim 50-60$\% of the $0.5-2$ keV emission from bright ($L_{\rm X} > 10^{38}$) late-type galaxies and  $\sim 40$\% of the $0.5-2$ keV emission from bright early-type galaxies for our best fitting model (205).  Hence, we find that hot gas emission has an appreciable effect on the galaxy XLFs.   See section 4 for more details on our results.
	
	\begin{deluxetable*}{cccccccccc}
\tablecolumns{8}
\tabletypesize{\scriptsize}
\tablewidth{0pt}
\tablecaption{Parameters and likelihood values for models referred to in this paper.  A full list is available online.}
\tablehead{ \colhead{Model} &
	\colhead{$\alpha_{\rm CE}$ \tablenotemark{a}} &
	 \colhead{IMF exponent} &
	 \colhead{$\eta_{\rm wind}$ \tablenotemark{b}} &
	\colhead{CE-HG \tablenotemark{c}} &
	\colhead{q distribution \tablenotemark{d}} &
	\colhead{DC BH kick \tablenotemark{e}} &
	\colhead{rank \tablenotemark{f}} &
	\colhead{$Log(L(O|M)/L_{ref})$ \tablenotemark{g}}
	}
\startdata

	205 & 0.1 & 2.7 & 2.0 & No & 50-50 & 0.0 & 1 & 0.0000000\\
	229 & 0.1 & 2.7 & 2.0 & Yes & 50-50 & 0.0 & 2 & -0.057250977\\
	277 & 0.1 & 2.7 & 2.0 & Yes & 50-50 & 0.1 & 3 & -0.53945923\\
	245 & 0.1 & 2.7 & 1.0 & No & 50-50 & 0.1 & 4 & -1.6570358\\
	253 & 0.1 & 2.7 & 2.0 & No & 50-50 & 0.1 & 5 & -1.7356873\\
	273 & 0.1 & 2.35 & 2.0 & Yes & 50-50 & 0.1 & 6 & -2.3401947\\
	269 & 0.1 & 2.7 & 1.0 & Yes & 50-50 &  0.1 & 12 & -5.6114807\\
	249 & 0.1 & 2.35 & 2.0 & No & 50-50 & 0.1 & 10 & -4.1068573\\
	248 & 0.5 & 2.7 & 1.0 & No & 50-50 & 0.1 & 55 & -47.292496\\
	197 & 0.1 & 2.7 & 1.0 & No & 50-50 & 0.0 & 50 & -42.623398\\
	241 & 0.1 & 2.35 & 1.0 & No & 50-50 & 0.1 & 59 & -51.771774\\
	261 & 0.1 & 2.7 & 0.25 & No & 50-50 & 0.1 & 81 & -92.800415\\
	53 & 0.1 & 2.7 & 1.0 & No & Flat & 0.1 & 22 & -15.239967\\
\enddata

\tablenotetext{a}{CE efficiency parameter}
\tablenotetext{b}{Stellar wind strength parameter}
\tablenotetext{c}{0: CE from Hertzsprung gap donor allowed, 1: not allowed}
\tablenotetext{d}{binary mass ratio distribution.}
\tablenotetext{e}{SN kicks given to direct collapse black holes.  0.0 = no SN kick given, 0.1 = small SN kick given}
\tablenotetext{f}{The rank of the model based on the likelihood value}
\tablenotetext{g}{Log of the ratio of the likelihood of the given model to that of the highest likelihood model}
\end{deluxetable*}
	
	\begin{figure*}
	\centering
	\includegraphics[scale=0.75]{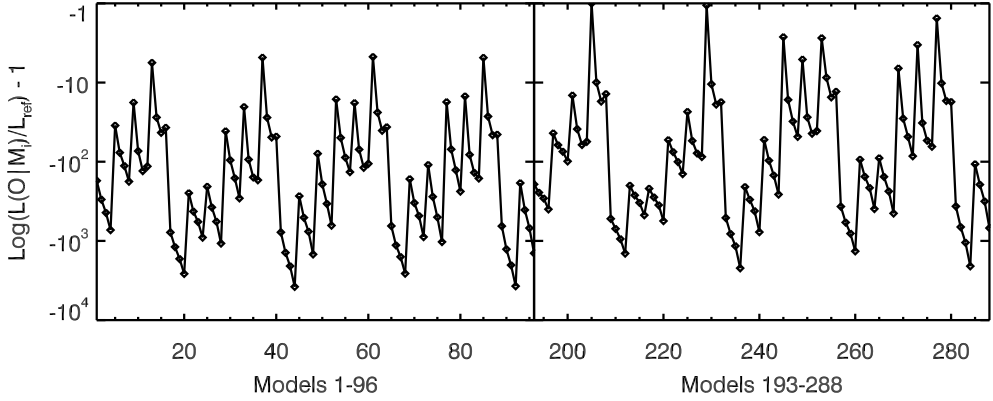}
	\caption{\label{Likelihood Values}Likelihood values for all of the models used in this work.  $L_{ref}$ is the highest likelihood value among our 192 models.  The model numbers correspond to the same models used in F12, though here we exclude models 97-192, as F12 found them to be inconsistent with observations.}
	\end{figure*}

\subsection{Statistical Analysis}

From deep \textit{Chandra} survey observations, T\&G08 present early-
and late-type galaxy counts in several luminosity bins and two
redshift intervals, $0 \leq z < 0.4$ and $0.4 \leq z < 1.4$.  They also provide total galaxy counts split into three redshift intervals, but for our analysis we will focus on the early and late type counts.
Associated with each count is a survey volume that depends on both
luminosity and redshift.  Let the set of counts be $d = \{d_{i,j} | i
= 1 \ldots N, j = 1 \ldots M \}$ and associated volumes be $V_{i,j}$,
where $i$ ranges over the $N=5$ luminosity bins and $j$ ranges over
the $M=2$ redshift bins.  T\&G08 assume that the $d_{i,j}$ are subject
to Poisson statistical errors.  

Similarly, for a particular choice of parameters, $\theta$, (for
example, see Table~\ref{Likelihood Values}) our model produces a set
of counts $n = \{n_{i,j} | i = 1 \ldots N, j = 1\ldots M\}$ for
galaxies in each luminosity and redshift bin in our $(100 \mathrm{Mpc}
/ h)^3$ model volume.  We assume that the counts $n$ are drawn from a
Poisson distribution with (unknown) means $\lambda = \{\lambda_{i,j} |
i = 1 \ldots N, j = 1 \ldots M\}$ (note:  this $\lambda$ is separate from the one used before to describe the CE efficiency parameter).  Because we only observe one
particular set of counts, $n$, we do not measure the rates $\lambda$
implied by our model directly, but instead must treat $\lambda$ as a
nuisance parameter whose distribution under the observed $n$ must be
integrated over.

Bayes' rule relates the posterior probability of model parameters
$\theta$, $p(\theta | d)$, to the likelihood of the data under the
model, $p(d | \theta)$, the prior probability of the model parameters
before the data have been observed, $p(\theta)$, and a normalizing
constant, $p(d)$, called the evidence, that is independent of $\theta$
via
\begin{equation}
  p(\theta | d) = \frac{p(d|\theta) p(\theta)}{p(d)}.
\end{equation}
Writing the likelihood in terms of the (unknown) true mean $\lambda$
implied by the model, we have 
\begin{equation}
  p(d|\theta) = \prod_{i,j} \int d\lambda_{i,j} \, p\left(d_{i,j} |
  \lambda_{i,j}\right) p\left(\lambda_{i,j} | \theta\right),
\end{equation}
where 
\begin{equation}
  \label{eq:likelihood-lambda}
  p\left(d_{i,j} | \lambda_{i,j} \right) =
  \frac{\left(v_{i,j}\lambda_{i,j}\right)^{d_{i,j}}}{d_{i,j}!}
  \exp\left( - v_{i,j} \lambda_{i,j} \right),
\end{equation}
is the Poisson probability of drawing $d_{i,j}$ counts in a volume
$V_{i,j} = v_{i,j} \left( 100 \mathrm{Mpc} / h \right)^3$ when the
underlying rate is $\lambda_{i,j}$ per $\left( 100 \mathrm{Mpc} / h
\right)^3$.  

The distribution of the underlying rates implied by our model,
$p\left(\lambda_{i,j}|\theta\right)$ must be estimated from the
observed $n_{i,j}$.  Applying Bayes' rule again, we have 
\begin{equation}
  p\left( \lambda_{i,j} | \theta \right) = p\left(\lambda_{i,j} |
  n_{i,j}(\theta) \right) = \frac{p\left( n_{i,j} | \lambda_{i,j}
    \right) p\left( \lambda_{i,j} \right)}{p\left(n_{i,j} \right)}.
\end{equation}
The counts observed in a model with underlying rate $\lambda_{i,j}$
are Poisson distributed, so 
\begin{equation}
  p\left( n_{i,j} | \lambda_{i,j} \right) =
  \frac{\lambda_{i,j}^{n_{i,j}}}{n_{i,j}!} \exp\left( - \lambda_{i,j}
  \right).
\end{equation}
We choose a Jeffreys prior%
\footnote{Note that the use of the Jeffreys prior implies that $\left
  \langle \lambda_{i,j} \right \rangle = n_{i,j} + \frac{1}{2}$.  A
  flat prior would have $\left \langle \lambda_{i,j} \right \rangle =
  n_{i,j} + 1$.  Both of these priors produce well-defined likelihoods
  even when $n_{i,j} = 0$ with $d_{i,j} \neq 0$. The
  maximum-likelihood estimator, $p\left( \lambda_{i,j} | \theta
  \right) = \delta\left( \lambda_{i,j} - n_{i,j} \right)$, while
  unbiased, produces likelihoods of zero if $n_{i,j} = 0$, even if
  only a single count appears in that bin of the data (i.e.\ $d_{i,j}
  = 1$).  A prior that gives $\left \langle \lambda_{i,j} \right
  \rangle = n_{i,j}$ (i.e.\ a prior that gives a distribution with
  unbiased mean) is $p\left(\lambda_{i,j}\right) =
  \lambda_{i,j}^{-1}$, which results in a non-normalizable likelihood
  when $n_{i,j} = 0$.} %
on $\lambda_{i,j}$,
\begin{equation}
  p\left( \lambda_{i,j} \right) = \frac{1}{\sqrt{\lambda_{i,j}}},
\end{equation}
whence
\begin{equation}
  \label{eq:p-lambda}
  p\left( \lambda_{i,j} | \theta \right) =
  \frac{\lambda_{i,j}^{n_{i,j}-\frac{1}{2}}}{\Gamma\left( n_{i,j} +
    \frac{1}{2} \right)} \exp\left( - \lambda_{i,j} \right).
\end{equation}

Combining Eq.~\eqref{eq:p-lambda} and
Eq.~\eqref{eq:likelihood-lambda}, we find that 
\begin{equation}
  \label{eq:likelihood}
  p(d | \theta) = \prod_{i,j}
  \frac{v_{i,j}^{d_{i,j}}\Gamma\left(\frac{1}{2} + d_{i,j} +
    n_{i,j}\right)}{\left(1+v_{i,j}\right)^{\frac{1}{2} + d_{i,j} +
      n_{i,j}} d_{i,j}! \Gamma\left( \frac{1}{2} + n_{i,j} \right)}
\end{equation}
We choose a flat prior on the model parameters, $\theta$, so that the
posterior is proportional to the likelihood in
Eq.~\eqref{eq:likelihood}:
\begin{equation}
  p(\theta | d) \propto p(d | \theta).
\end{equation}
	
        \begin{figure*}
	\centering
	\includegraphics[scale=0.7]{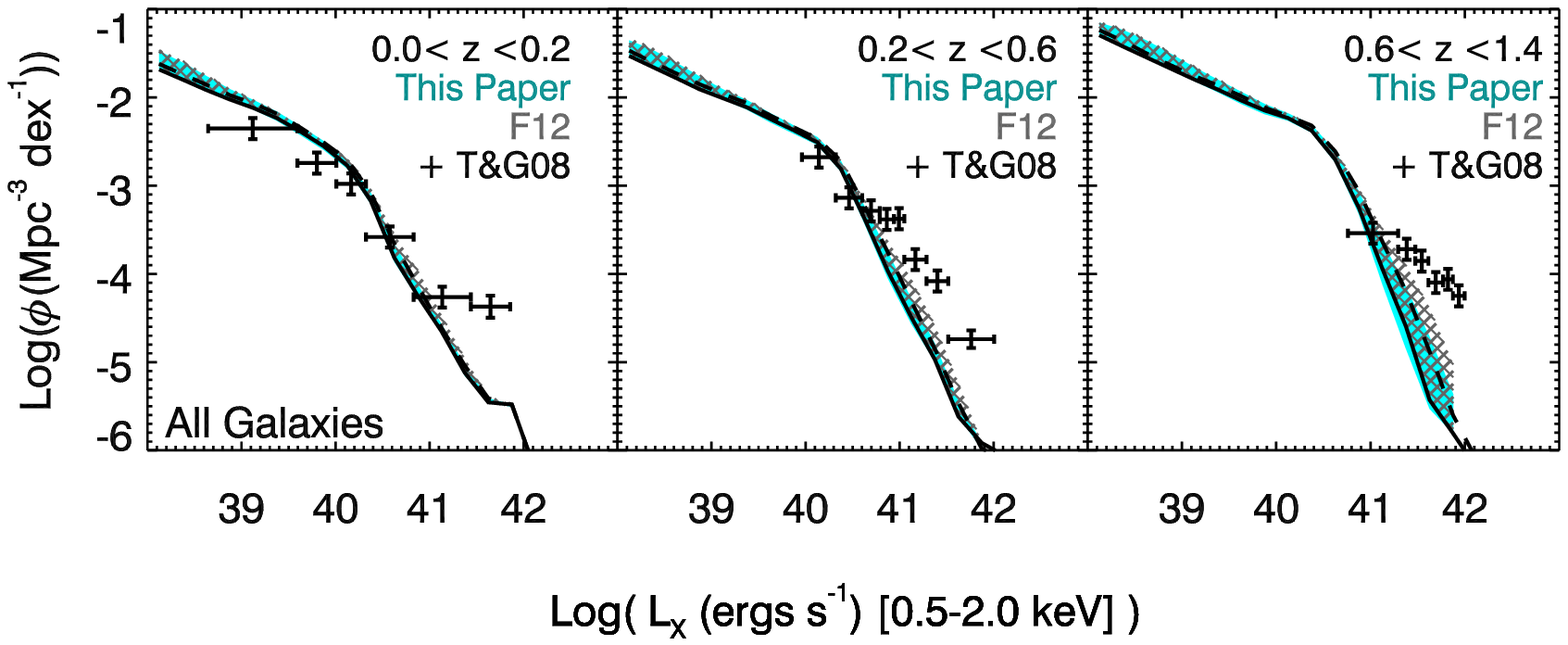}
	\includegraphics[scale=0.5]{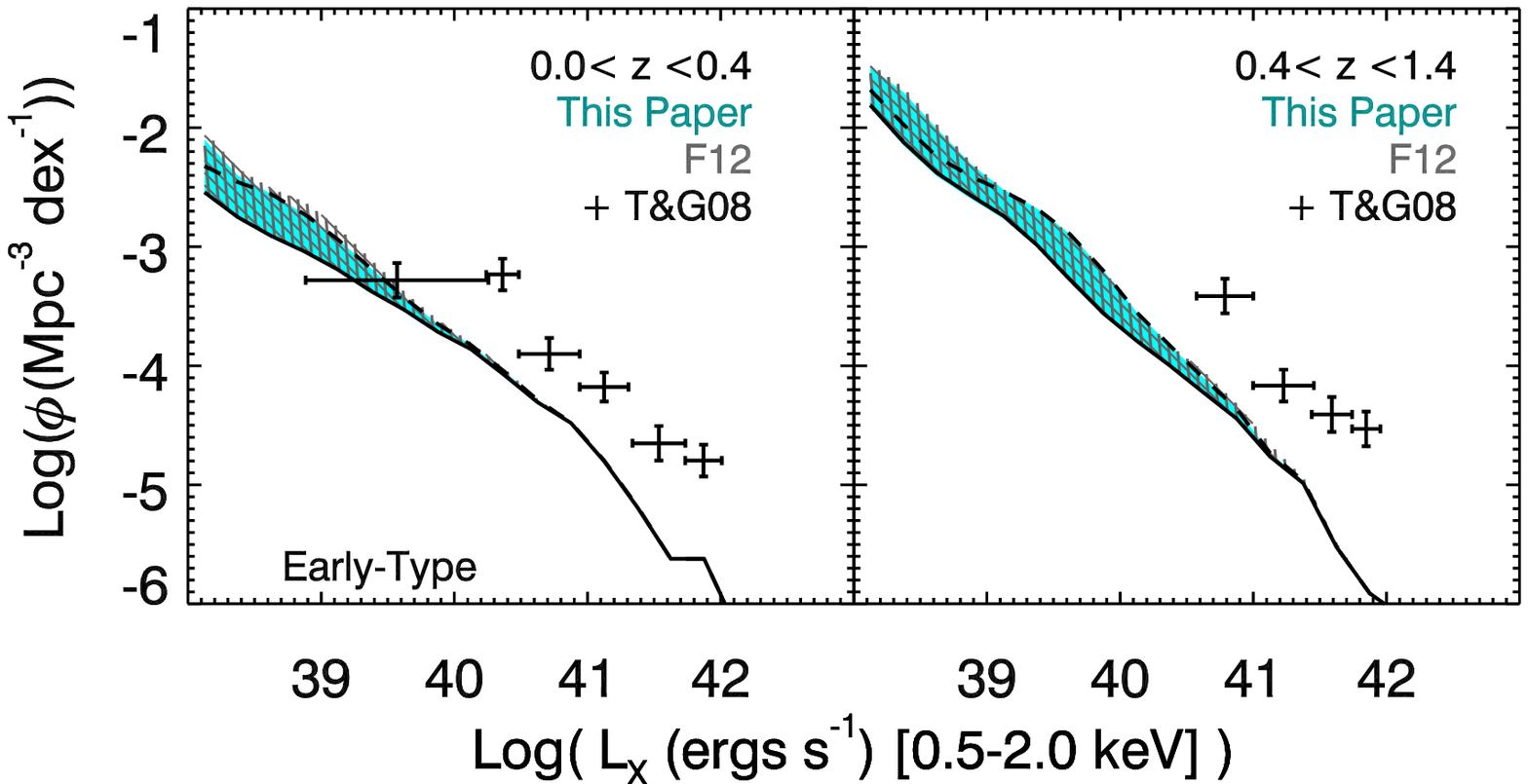}
	\includegraphics[scale=0.5]{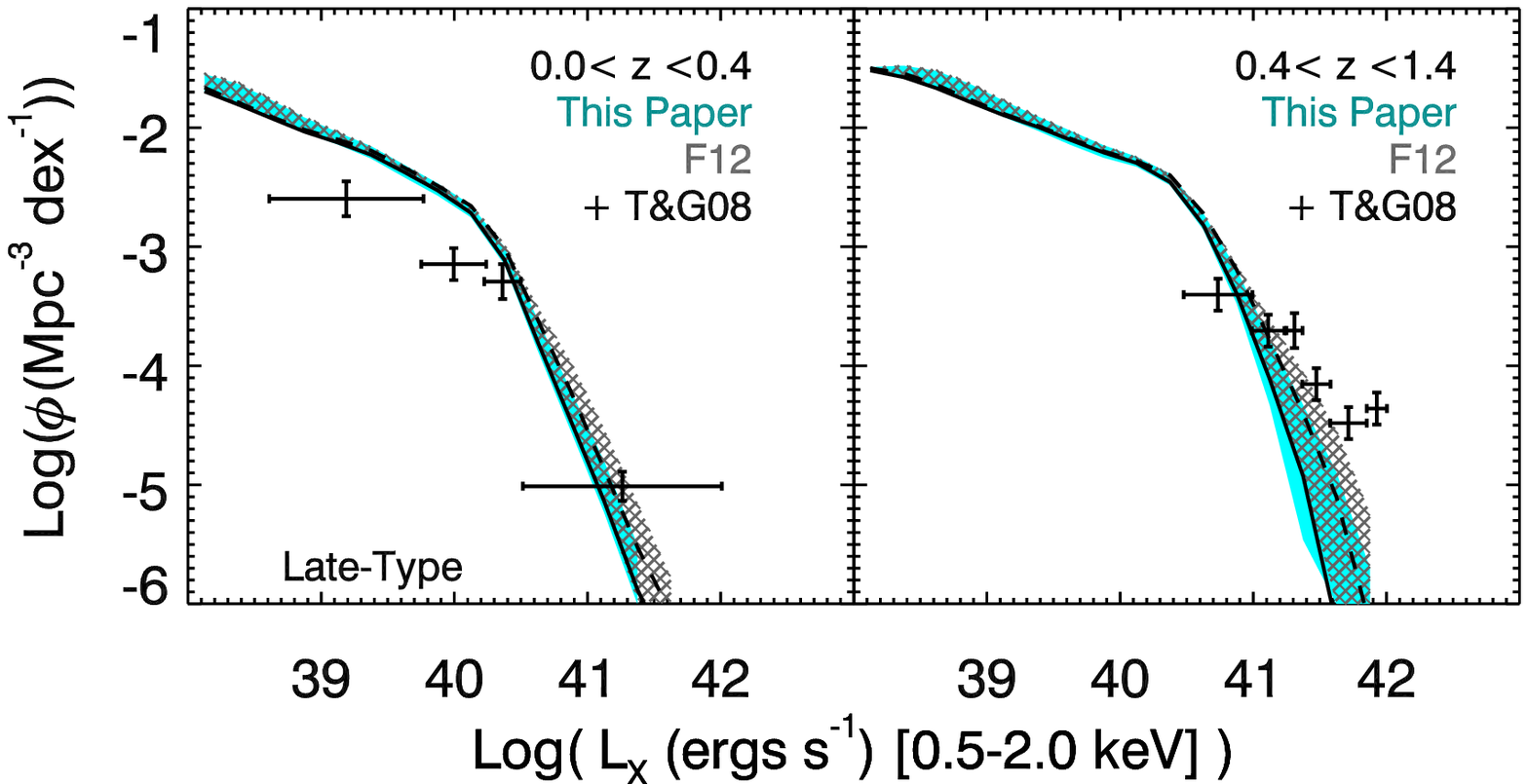}
	\caption{The cyan region shows the area bounded by the six highest likelihood models from this study, the grey checkered region shows that bounded by the six highest likelihood models from F12, the black points with error bars are data from T\&G08, the solid black lines are from our highest likelihood model (205), and the dashed black lines are from the highest likelihood model from F12 (245).  Top:  Total galaxy population.  Bottom left:  Early-type galaxies.  Bottom right:  Late-type galaxies. The XLFs from our highest likelihood models are very similar to those from the F12 models, however they underproduce bright early-type galaxies and very bright ($L_{\rm X} > 10^{41} ergs~s^{-1}$) late-type galaxies compared with observations.  Section 4.1 discusses the causes of these discrepancies in more detail.}
	\end{figure*}
	
	\begin{figure}
	\centering
	\includegraphics[scale=0.5]{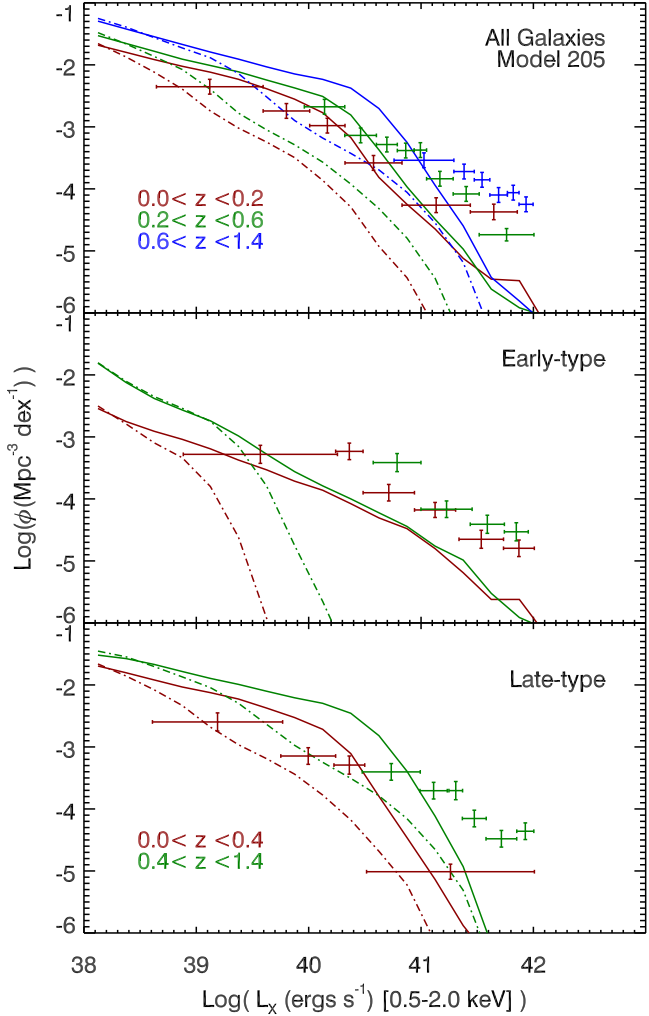}
	\caption{ Luminosity function for the total galaxy population using our highest likelihood model, number 205. Solid lines show the XLFs from our models with both XRB and hot gas emission.  The dot-dashed lines show XLFs for just XRB emission.  Data points and associated error bars are taken from T\&G08.  Consistent with the analysis of T\&G08, the overall XLF evolution is driven almost entirely by late-type galaxies.  However, the model fails to reproduce the correct normalization of the Early-type XLF and the shape of the observed XLF of very bright ($L_{\rm X} > 10^{41} ergs~s^{-1}$) late-type galaxies.  Section 4.1 discusses the possible causes for these discrepancies in more detail. Hot gas plays an important role in the normalization of the XLF, especially for early-type galaxies, where the hot gas contribution also affects the XLF evolution with redshift.}
	\end{figure}
	
	\begin{figure}
	\centering
	\includegraphics[scale=0.5]{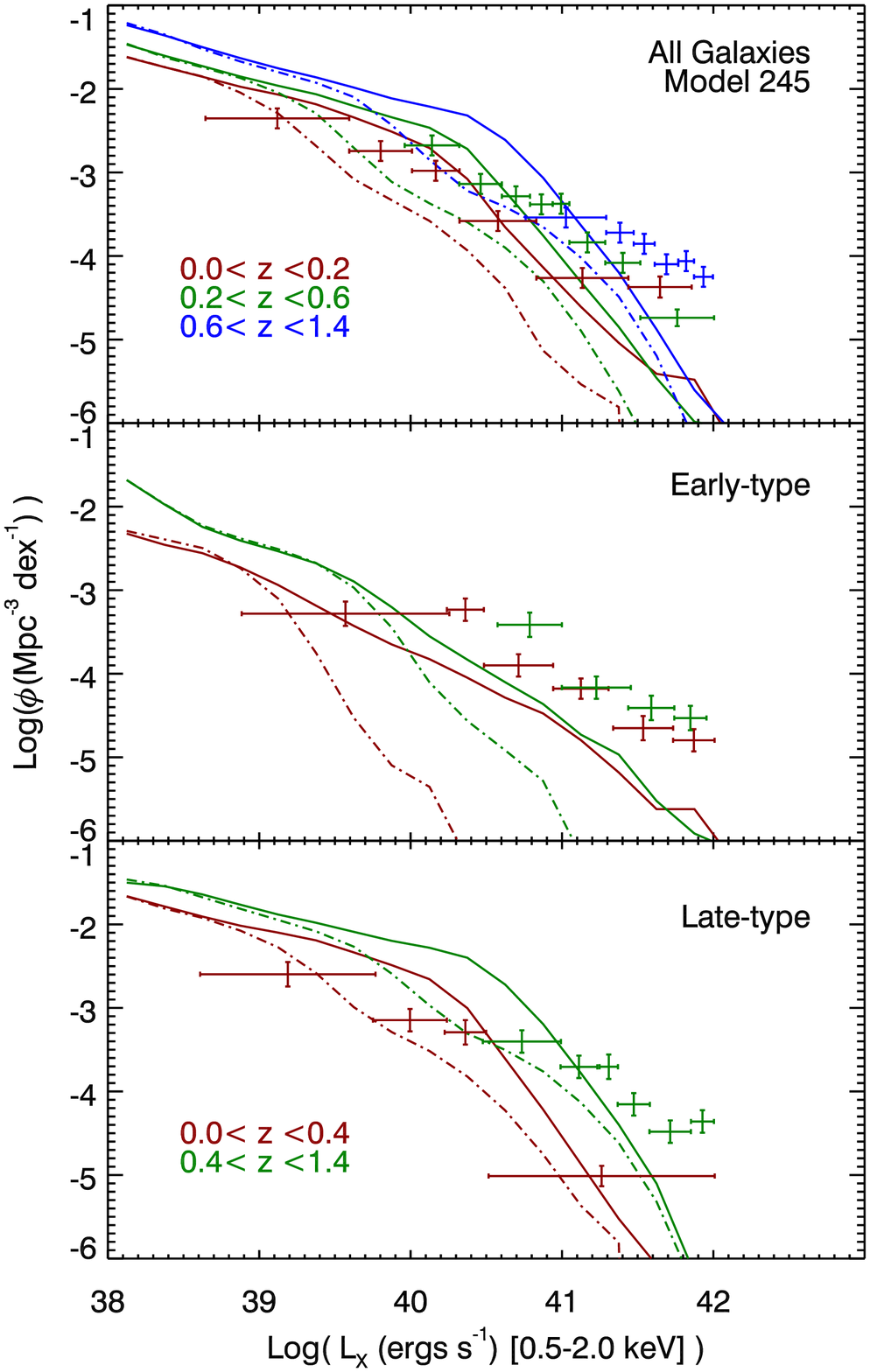}
	\caption{Same as figure 4, but for model 245, the highest likelihood model from F12 and the fifth highest likelihood models from this work.  These XLFs are similar to that of our highest likelihood model, 205.}
	\end{figure}
	
\section{Results and Discussion}
	
	Figure 2 plots the likelihood values for each model used in this study and shows that likelihood values are very sensitive to model parameters and that only a few models are able to accurately reproduce the observed XLFs.  Therefore, this comparison is very useful for eliminating regions of our model parameter space. Table 2 lists the six models with $Log(L(O|M)/L_{ref}) > -3$, where $L_{ref}$ is the highest likelihood value among our 192 models.  These models are 205, 229, 277, 245, 253, and 273.  These models all have low CE efficiencies ($\alpha_{CE} = 0.1$), a 50-50 mass ratio distribution, an IMF exponent of $-2.7$ (with the exception of model 273), and $\eta_{\rm wind} = 1.0-2.0$.  Recall that our CE efficiency parameter really represents $\alpha_{\rm CE} \times \lambda$, so low values of $\alpha_{CE}$ could alternatively be interpreted as these systems having a high envelope binding energy, which has been found to be true for massive stars \citep{2012arXiv1202.4901D}. The models that have $\eta_{\rm wind} = 2.0$, $\alpha_{CE} = 0.1$, IMF exponent of $-2.7$, and a flat q-distribution also have fairly high likelihood values (model numbers 13, 37, 61, and 85).
	
	It should be noted that our likelihood calculation takes into account the number of samples in each bin.  The overall likelihood values are much more sensitive to bins with higher sample counts (i.e. those at lower luminosity).
	
	Figure 2 and Table 2 show that allowing/not allowing CE-HG phases has  only a relatively small effect on the likelihoods of our best models.  In addition, DC BH kicks only have an appreciable effect on likelihoods for models with lower wind mass loss rates ($\eta_{\rm wind} = 0.25-1.0$).  For models with $\eta_{\rm wind} = 2.0$, such as those that make up the majority of our top models, DC BH kicks have little effect.  Thus, these two parameters are not very well constrained by our analysis.
				
	\subsection{Comparison with \citet{Fragos2012} and \citet{2008A&A...480..663T}}
	
	F12 use the same PS models used in this work (with the inclusion of pure ``twins'' models) to study the evolution of the overall population of XRBs in the Universe.  They compare with X-ray observations of local galaxies that give estimates of the specific X-ray luminosity of XRBs in the local Universe \citep{2010ApJ...724..559L,2011ApJ...729...12B,2012MNRAS.419.2095M}.  They calculate the likelihood of each model based on these data and find the six highest likelihood models to be, in order of likelihood, 245, 229, 269, 205, 249, and 273 (see table 2 for model parameters and likelihood values from this study).  Figure 3 compares the six highest likelihood models from this work with those of F12.  Four out of our top six models are also among the six highest likelihood models from F12, so it is no surprise that the region bounded by the models in this work is very similar to that bounded by the models from F12.
	
	Figure 4 shows XLFs for our highest likelihood model (205), with and without hot gas emission, plotted against the data from T\&G08.  Figure 5 plots XLFs similar to figure 4, but for the highest likelihood model from F12 (245).  These plots show that our models are able to reproduce the redshift evolution of the observed XLFs. Consistent with the analysis of T\&G08, the XLF evolution is driven almost entirely by late-type galaxies.  
Our models also reproduce the shape of the early-type XLF and the normalization of the late-type XLF, though they drastically underproduce bright early-type galaxies and they fail to reproduce the shape of the bright ($L_{\rm X} > 10^{41} ergs~s^{-1}$) end of the late-type XLF.   
	
	Figures 4 and 5 also show that hot gas can have a large effect on the shape and normalization of the XLF, showing that hot gas emission dominates the integrated X-ray luminosity of the brightest galaxies in our sample.  Adding in the hot gas emission suppresses the redshift evolution for the early-type galaxy XLF. For galaxies with $L_x > 10^{40}$ ergs/s at low ($z<0.8$) redshift, emission from XRBs accounts on average for only $1-5$\% and $\sim15$\% of early and late-type galaxy emission respectively.  However, as we will discuss in section 4.3, the XLFs are still rather sensitive to changes in our model parameters, as seen in the varying likelihood values shown in figure 2.  The important role of hot gas emission on the XLF means that our simplistic prescriptions for hot gas emission add a great deal of uncertainty to our models and could be a major source of the discrepancies between the models and observations, particularly for early-type galaxies.   While our method is motivated by observations, it does not take into account the internal characteristics of the gas that contribute to its emission, such as density and metallicity.  Further, the relations used for early-type galaxies were derived only from low redshift sources, which may not be accurate for the high redshift galaxies studied here.
	
	In addition, the G11 semi-analytic model underproduces massive galaxies at high redshift.  This will lead to less large elliptical galaxies, which could explain part of our discrepancy at higher redshift.
	
	Another aspect of our models that can account for the underproduction of bright early-type galaxies is that our PS models only take into account LMXBs formed in the field and not those formed dynamically in globular clusters. Dynamically formed LMXBs are believed to play a significant role in old, massive, GC-rich elliptical galaxies. \citep[e.g.][]{2008ApJ...689..983H,2012arXiv1202.2331Z}.  These LMXB populations can have a significant contribution to the integrated X-ray luminosity of bright early-type galaxies, as they can make up over half of the total number of LMXBs in a galaxy \citep{Irwin2005}.  So, including dynamically formed LMXBs in our models could increase the number of bright LMXBs, and therefore the total LMXB luminosity, in early-type galaxies by a factor of $\sim3$.  Changing the q-distribution in model 245 from a 50-50 to a flat distribution is a good proxy for this effect because it increases the LMXB population without changing the distribution of their physical properties (see section 4.3).  T12 show that doing this increases the total luminosity from LXRBs at all redshifts by a factor of 2.  We find that changing to a flat q-distribution increases the low luminosity end of the early-type XLF by  $\sim0.3$ dex (see figure 7).  The effect of including dynamically formed LMXBs could have a similar but greater effect, bringing the low luminosity end of the XLF closer to observations, but having little effect on higher luminosity galaxies.  A more detailed calculation will require information on the GC population of each galaxy, which is not included in the G11 catalog.
			
	For younger, star forming galaxies, our models also have only a very basic formula to simulate starburst activity, which can occur, e.g., due to galaxy mergers.  This would have a significant affect on the HMXB populations present in late-type galaxies, and the effect would not necessarily be constant with redshift, as merger rates may evolve with time \citep[e.g.][]{2006ApJ...638..686C}.  Thus, a more detailed SFH is needed to more accurately model HMXB populations of late-type galaxies.  
	
	In addition, the higher end of the observed late-type galaxy XLF is more at risk from AGN contamination, even with the efforts of T\&G08 to minimize this effect.  Since the observations of T\&G08, the depth of the X-ray data, combined with better multiwavelength data, have allowed for more accurate classifications of the X-ray sources.  Of the 56 1 Ms CDF-S sources used in T\&G08, we find 53 counterparts with  4 Ms exposure using a matching radius of 2.5 arcsec.  The missing three sources may have been false-positive sources in the 1Ms data.  Of these 53 sources, we find that 25 of them are classified as normal galaxies and 28 of them as AGN according to the six criteria highlighted in section 3.1 of \citet{2012ApJ...752...46L}.  Therefore, it is possible that the T\&G08 data points will be lowered by $\sim0.3$ dex.  However, it is difficult to know in detail how this affects the TG08 luminosity functions and recomputing the luminosity functions is beyond the scope of this work.
				
	\subsection{High Redshift Predictions}
	
	Figure 6 plots the X-ray luminosity density from normal galaxies as derived from our highest likelihood model, 205.  The overall evolution (black line in figure 6) is very similar to that of the observed SFH of the universe.  It is also similar to the evolution of the specific XRB X-ray luminosity of the universe predicted in F12, despite the inclusion here of hot gas emission.  This is evidence that XRBs drive the overall evolution of the normal galaxy X-ray luminosities out to at least $z=4$.  However, our predicted X-ray luminosity density reaches a maximum at $z\sim2.5$, which is lower compared with the XRB models in F12 that reach a maximum at $z\sim3$.  This can be attributed to the inclusion of hot gas emission in our models, which has already been shown to have a noticeable effect on the shape and evolution of our XLFs, particularly for early-type galaxies.  
	
	Splitting the galaxies into three luminosity bins, we find that the evolution of low ($10^{39} < L_{\rm X} < 10^{40} ergs~s^{-1}$) luminosity galaxy emission is small compared with the evolution for higher luminosity galaxies, which varies by an order of magnitude on the range $z=0$ to $z=4$.  The most luminous galaxies ($L_{\rm X} > 10^{41} ergs~s^{-1}$) reach a maximum around $z=3$, which also approximately corresponds to the time of maximum SFR density in the Universe.  Galaxies in the range $10^{40} < L_{\rm X} < 10^{41} ergs~s^{-1}$ reach a maximum around $z=2$.  In the local Universe, the low luminosity galaxies dominate the normal galaxy X-ray emission.  We do not go to higher redshift here because our hot gas emission prescription relies on galaxy morphology, which becomes harder to classify at higher redshifts \citep{2002PASP..114..797V}.

	\begin{figure}
	\centering
	\includegraphics[scale=0.5]{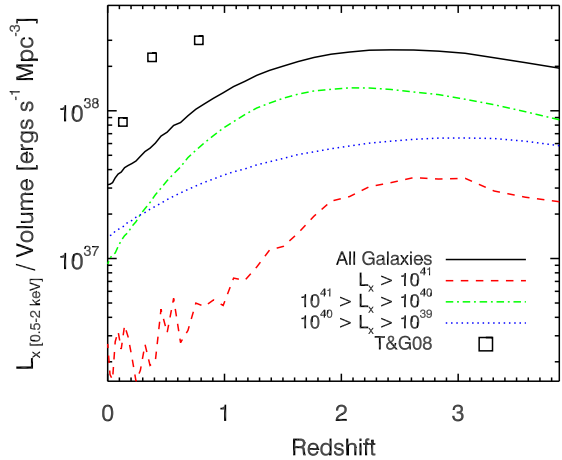}
	
	\caption{X-ray luminosity density of normal galaxies versus redshift for model 205.  The back line shows the contribution from all galaxies and the black squares show the total X-ray luminosity density from observations presented in T\&G08.  The X-ray luminosity density derived from our models is a factor of $2-3$ lower than the observations. This is due to the fact that our models, relative to observations, underproduce bright early-type galaxies and very bright ($L_{\rm X} > 10^{41}$) late-type galaxies.  Our model follows a similar evolution as the SFR density and the X-ray luminosity density from XRBs predicted in F12, indicating that XRBs continue to drive the evolution of the normal galaxy X-ray emission at higher redshifts.  However, our models reach a maximum luminosity density at $z\sim2.5$, compared with $z\sim3$ for the XRB models in F12.  This can be attributed to the inclusion of hot gas emission, which has been shown to have an appreciable effect the evolution of our XLFs when compared with XRB only models. The red, green, and blue lines plot the specific X-ray luminosity for different luminosity bins.  The amount X-ray emission from lower luminosity galaxies evolves much less with redshift than that from brighter galaxies.  In the local Universe, most of the normal galaxy X-ray emission comes from lower luminosity galaxies.}
	\end{figure}
	
	\begin{figure}
	\centering
	\includegraphics[scale=0.5]{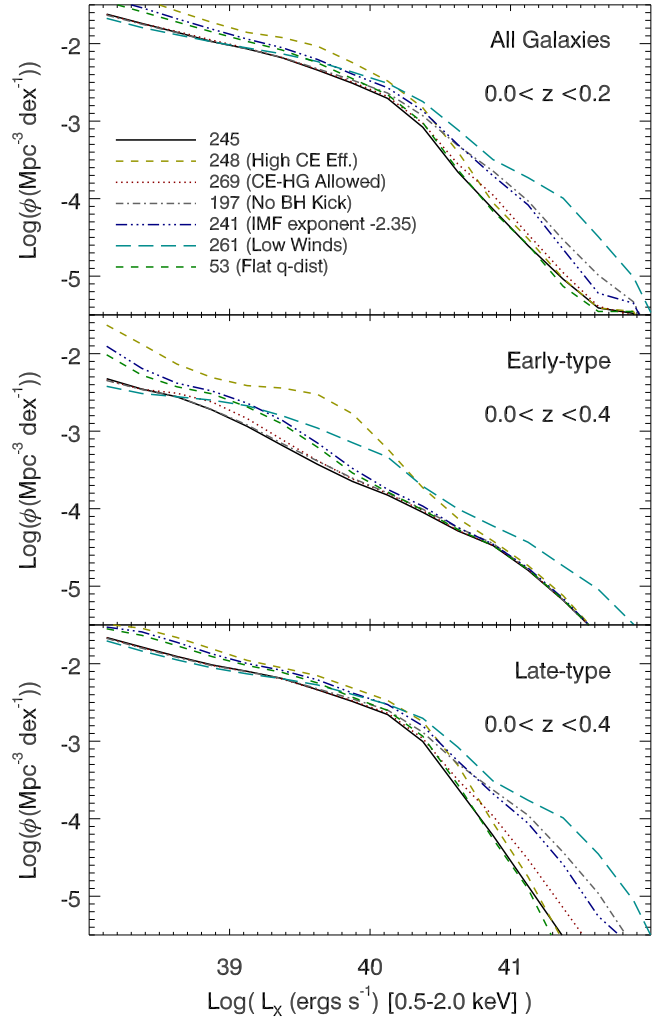}
	\caption{the highest likelihood model from F12(245, black line) compared with other models to illustrate the effects of different parameters on the shape of the XLF.  Different parameters have varying effects on the shape of the XLFs. Shown here are models with higher $\alpha_{CE}$ (248), CE-HG allowed (269), no BH kicks (197), steeper IMF (241), lower wind mass loss (261), and a flat distribution (53).}
	\end{figure}
	
	\subsection{Effects of parameters on XLFs}

	Figure 7 shows XLFs for different models compared to model 245.  Each model is chosen to encapsulate the effect that each parameter has on the shape of the XLF.  Several parameters have significant effects on the shape of the XLFs, while others have only minimal effects.  Model 245 was chosen because it has both a high likelihood and is more sensitive to certain changes in parameters, thus better illustrating the different effects on our XLFs.
	
	The common envelope efficiency parameter ($\alpha_{CE}$) dictates how efficiently orbital energy is converted to thermal energy that will expel the envelope.  A lower efficiency means that it will take more orbital energy to expel the envelope.  This parameter mainly effects LMXBs, as most LMXBs formed in the field must go through a CE phase.  The CE phase plays an important role in making the orbit close enough to allow for Roche lobe overflow (RLO), but a lower CE efficiency leads to even more orbital decay and a higher rate of mergers, overall decreasing the rate of LMXB formation.  This effect can be seen by comparing models 245 and 248.  HMXBs are not as strongly affected by changes in $\alpha_{CE}$, as they have other formation channels available that do not involve CE phases \citep{2010ApJ...725.1984L, 2010Natur.468...77V}.  This parameter mainly affects the lower luminosity end of the XLF, where a higher CE efficiency increases the number of bright galaxies due to an increased LMXB population.
	
	Wind mass loss rates effect the evolution of high mass stars in two major competing ways.  Higher wind mass loss rates will increase the accretion rates of wind-fed HMXBs, increasing their luminosity.  On the contrary, lower winds will result in a lower overall mass loss of the primary star and hence increase the formation rate of massive BHs.  BH-XRBs tend to be more luminous than XRBs with NS accretors.  This is because, on the one hand, they can form stable RLO XRB systems with massive companions and, on the other hand, BHs drive higher accretion rates due to their higher mass and, therefore, higher Eddington limits.  In this way, weaker stellar winds can increase the luminosities of both LMXB and HMXB populations.  Comparing models 245 and 261, we see that weaker stellar winds increases the number of bright early and late-type galaxies, so the latter effect is dominant.  While lower stellar winds will help our highest likelihood model match observations of early-type galaxies, it would overproduce bright late-type galaxies.

	Changing the initial binary mass ratio between a flat distribution and a 50\% twins and 50\% flat distribution effects both the HMXB and LMXB populations.  As stated earlier, the binary mass ratio effects the secondary star in the system, which will eventually become the donor star in most cases.  All XRBs are accreting mass onto a compact object, which can only come from a high mass progenitor.  Most LMXBs require a high initial mass ratio in order to ensure a high mass primary star that will evolve into a BH or NS with a lower mass companion.  The ``50-50'' distribution adopted here forces mass ratios close to 1 for half the binary population.  This will decrease the LMXB population while increasing the HMXB population, as HMXBs require ratios closer to 1.  Changing from ``50-50'' to a flat distribution (comparing models 245 and 53) increases the lower luminosity end of the early and late-type galaxy XLFs and slightly decreases the number of very bright late-type galaxies galaxies.
	
	Allowing small natal kicks for direct collapse BHs affects the HMXB and LMXB population in two competing ways.  On the one hand, natal kicks can enhance the formation of RLO-HMXBs \citep{2010ApJ...725.1984L}.  On the other hand, natal kicks inject energy into the binary system and could result in the widening or the complete disruption of the system, thereby decreasing the formation HMXB and LMXB with BH accretors.  \citet{2009ApJ...699.1573L} find that imparting small natal kicks to DC BHs is necessary in order to reproduce the lack of observed wide orbit BH-XRBs.  Comparing models 245 and 197, not allowing natal kicks increases the high luminosity end of the late-type galaxy XLF and has little effect on the early-type galaxy population.
	
	Within our grid of PS models, we also have the IMF power law exponent as a free parameter, allowing it to be either $-2.35$ or $-2.7$.   It is instructive to note that the IMF referred to in this work represents the integrated galaxy IMF \citep{2005ApJ...625..754W, 2003ApJ...598.1076K} and that this work only probes the high mass region of the IMF, since the primary stars that are created via sampling the IMF must be massive enough to form a BH or NS.  The slope of the power law at the high-mass end of the IMF affects the population of XRBs in a way similar to stellar winds, in the sense that a flatter IMF will produce relatively more massive BHs compared to a steeper one.  A flatter IMF will result in more bright LMXBs and HMXBs.  This affect can be seen by comparing models 245 and 241.  As expected, a flatter IMF results in a higher number of bright galaxies, though the very high luminosity end of the early-type galaxy XLF is not strongly affected by this parameter.  This is not surprising, as figure 4 shows that hot gas emission dominates the high luminosity end of the XLF for early-type galaxies.
	
	Finally, we found that allowing or not for all possible outcomes in CE phases with donor stars in the Hertzsprung gap has only a small affect on the galaxy XLFs, slightly increasing the number of bright late-type galaxies (compare models 245 and 269).  Thus, although this parameter affects the shape of the XLF of individual XRBs in a galaxy \citep{2012ApJ...749..130L}, it has a negligible effect on the integrated X-ray luminosity of a galaxy.
	
	The models that we find to agree best with observations have direct collapse BH natal kicks, low CE efficiency, steep IMFs, $\eta_{wind} = 1.0-2.0$, and a 50-50 mass ratio distribution.  As outlined above, these parameters make for a limited population of both LMXBs and BH-XRBs.  The HMXB population is limited by the steeper IMF, BH natal kicks, and higher winds, but also strengthened by the 50-50 mass ratio distribution.  However, the latter effect on the XLF is much weaker, as shown in figure 7.
	
	\section{Conclusions}
	
	Using data from the \textit{Millennium Cosmological Simulation} and the semi-analytical analysis conducted by G11 in tandem with the binary population synthesis code, \textit{StarTrack}, we simulated the population of XRBs within normal galaxies in a large volume of the Universe from $z = 0$ to $\sim 20$.  Assuming that galaxy X-ray emission is solely due to XRBs and hot gas, we calculated the integrated X-ray luminosity of each galaxy in this cosmic volume and compared the resulting galaxy XLFs to the observational XLFs of T\&G08.
	
	In this paper, we presented data from 192 binary population synthesis models, varying parameters that have the largest effect on binary star evolution (see Table 1 for a list of the parameters).  We use a likelihood calculation method to compare  each model with the results from T\&G08.  From this analysis we find that our theoretical XLFs are sensitive to many of our model parameters and that only a few of our models are able to reproduce the most recent observations of X-ray bright normal galaxies.  Our highest likelihood models are also among the highest likelihood models from a separate analysis presented in F12.  This confirms that our results are consistent with their separate analysis, which compares these same models with the observed overall emission from LMXB and HMXB populations in the local Universe.  To have only $\sim10$ models from our 192 model grid best match observations in two separate analysis shows that we are able to provide self consistent constraints on the XRB parameter space.
		
	We find that our highest likelihood models are those with a lower LMXB population due to a low CE efficiency and a 50-50 mass ratio distribution, and a lower BH-XRB and HMXB population due to higher winds,  and a steeper IMF.  Our models do well in reproducing the normalization and evolution of the total and late-type galaxy XLFs, as well as the evolution and shape of the early-type XLF.  
		
	Our models show that hot gas emission has a large effect on the shape of the XLFs, and it significantly affects the redshift evolution of the early-type galaxy XLF, causing it to remain nearly constant out to $z=1.4$.
		
	We show that the observed redshift evolution of the normal galaxy XLF continues out to higher redshift, with the specific normal galaxy X-ray luminosity evolving in a way similar to the SFH of the universe and consistent with the evolution of XRB emission found in F12.  This is evidence that the XLF evolution is driven by XRB evolution even out to higher redshifts.  Our models also show that hot gas emission causes the point of maximum normal galaxy X-ray luminosity density to shift to lower redshift compared with the XRB models in F12.
			
	However, despite these many successes, our models do not perfectly reproduce the observed XLFs.  In particular, they fail to reproduce the observed normalization of the early-type galaxy XLF, greatly underestimating the number of bright early-type galaxies.  Our highest likelihood models also fail to reproduce the shape of the high ($ L_{\rm X}>10^{41} ergs~s^{-1}$) luminosity end of the late-type galaxy XLF, particularly for higher redshifts.  
	
	Our models have limitations that may have caused these discrepancies.  For one, we do not take into account dynamically formed LMXBs, which could significantly increase the normalization of the model early-type galaxy XLF.  For late-type galaxies, the XRB luminosities have a higher contribution from HMXB populations, which are very sensitive to evolving SFHs. However, the SFHs used from the G11 catalog are limited in their detail and our method for simulating the effect of starbursts is very rudimentary.  In addition, our prescription for hot gas, though based on observations, is very basic and could add inaccuracy to our X-ray luminosities as well as the selection of our best fitting models.  A more detailed model is needed to more accurately model the hot gas emission, particularly in early-type galaxies.
	
	In addition to limitations in our models, the observations of very bright galaxies are subject to the possibility of AGN contamination, which could artificially increase the observed high luminosity data points from T\&G08.
				
	Despite these shortcomings, this work represents a first careful attempt to study how XRBs control the Lx distributions of different types of galaxies. As such, it provides an important theoretical base for future X-ray observations of normal galaxies at high redshift.  It also shows that XRB populations are closely linked with the growth of galaxies.  This work lays the ground for future work using X-ray observations and cosmological simulations of galaxies to provide a new way to constrain our models of binary evolution, as well as study the role played by XRBs in galaxy formation and evolution through feedback processes.
	\\
	\\
	\\
	The authors thank the anonymous referee whose comments and suggestions have helped to improve this paper.  TF acknowledges support from the CfA and the ITC prize fellowship programs.  B.D.L. thanks the Einstein Fellowship program.  PT acknowledges support through a NASA Post-doctoral Program Fellowship at Goddard Space Flight Center, administered by Oak Ridge Associated Universities through a contract with NASA. Resources supporting this work were provided by the  Northwestern University Quest High Performance Computing (HPC) cluster and by the NASA High-End Computing (HEC) Program through the NASA Center for Climate Simulation (NCCS) at Goddard Space Flight Center.  KB acknowledges support from MSHE grant N203 404939.  VK acknowledges support for this work from NASA ADP grant NNX12AL39G (sub-contract to Northwestern U.)
	
	\bibliographystyle{apj}

\appendix
 
 \section{To appear as supplemental online-only material}
 \LongTables
 	\begin{deluxetable}{cccccccccc}
	\centering
\tablecolumns{8}
\tabletypesize{\scriptsize}
\tablewidth{0pt}
\tablecaption{A complete list of all PS models used in this work.}
\tablehead{ \colhead{Model} &
	\colhead{$\alpha_{\rm CE}$ \tablenotemark{a}} &
	 \colhead{IMF exponent} &
	 \colhead{$\eta_{\rm wind}$ \tablenotemark{b}} &
	\colhead{CE-HG \tablenotemark{c}} &
	\colhead{q distribution \tablenotemark{d}} &
	\colhead{DC BH kick \tablenotemark{e}} &
	\colhead{rank \tablenotemark{f}} &
	\colhead{$Log(L(O|M)/L_{ref})$ \tablenotemark{g}}
	}
\startdata
1 & 0.1 & 2.35 & 1.0 & No & Flat & 0.0 & 105 & -172.65289\\
2 & 0.2 & 2.35 & 1.0 & No & Flat & 0.0 & 127 & -298.51453\\
3 & 0.3 & 2.35 & 1.0 & No & Flat & 0.0 & 142 & -439.21210\\
4 & 0.5 & 2.35 & 1.0 & No & Flat & 0.0 & 161 & -727.54431\\
5 & 0.1 & 2.7 & 1.0 & No & Flat & 0.0 & 42 & -33.895195\\
6 & 0.2 & 2.7 & 1.0 & No & Flat & 0.0 & 72 & -75.746765\\
7 & 0.3 & 2.7 & 1.0 & No & Flat & 0.0 & 89 & -111.71681\\
8 & 0.5 & 2.7 & 1.0 & No & Flat & 0.0 & 106 & -176.81656\\
9 & 0.1 & 2.35 & 2.0 & No & Flat & 0.0 & 29 & -16.885155\\
10 & 0.2 & 2.35 & 2.0 & No & Flat & 0.0 & 69 & -72.331543\\
11 & 0.3 & 2.35 & 2.0 & No & Flat & 0.0 & 93 & -129.83803\\
12 & 0.5 & 2.35 & 2.0 & No & Flat & 0.0 & 90 & -113.39149\\
13 & 0.1 & 2.7 & 2.0 & No & Flat & 0.0 & 11 & -4.6178589\\
14 & 0.2 & 2.7 & 2.0 & No & Flat & 0.0 & 37 & -26.552711\\
15 & 0.3 & 2.7 & 2.0 & No & Flat & 0.0 & 49 & -41.962570\\
16 & 0.5 & 2.7 & 2.0 & No & Flat & 0.0 & 44 & -36.056236\\
17 & 0.1 & 2.35 & 0.25 & No & Flat & 0.0 & 165 & -779.69275\\
18 & 0.2 & 2.35 & 0.25 & No & Flat & 0.0 & 176 & -1189.3790\\
19 & 0.3 & 2.35 & 0.25 & No & Flat & 0.0 & 184 & -1683.3291\\
20 & 0.5 & 2.35 & 0.25 & No & Flat & 0.0 & 190 & -2598.7773\\
21 & 0.1 & 2.7 & 0.25 & No & Flat & 0.0 & 119 & -247.78528\\
22 & 0.2 & 2.7 & 0.25 & No & Flat & 0.0 & 141 & -423.04449\\
23 & 0.3 & 2.7 & 0.25 & No & Flat & 0.0 & 152 & -572.43054\\
24 & 0.5 & 2.7 & 0.25 & No & Flat & 0.0 & 169 & -904.26648\\
25 & 0.1 & 2.35 & 1.0 & Yes & Flat & 0.0 & 112 & -204.92599\\
26 & 0.2 & 2.35 & 1.0 & Yes & Flat & 0.0 & 136 & -375.42163\\
27 & 0.3 & 2.35 & 1.0 & Yes & Flat & 0.0 & 151 & -568.51801\\
28 & 0.5 & 2.35 & 1.0 & Yes & Flat & 0.0 & 173 & -1074.5802\\
29 & 0.1 & 2.7 & 1.0 & Yes & Flat & 0.0 & 48 & -40.053345\\
30 & 0.2 & 2.7 & 1.0 & Yes & Flat & 0.0 & 82 & -94.520035\\
31 & 0.3 & 2.7 & 1.0 & Yes & Flat & 0.0 & 101 & -160.10526\\
32 & 0.5 & 2.7 & 1.0 & Yes & Flat & 0.0 & 125 & -288.00360\\
33 & 0.1 & 2.35 & 2.0 & Yes & Flat & 0.0 & 32 & -19.284058\\
34 & 0.2 & 2.35 & 2.0 & Yes & Flat & 0.0 & 80 & -92.763794\\
35 & 0.3 & 2.35 & 2.0 & Yes & Flat & 0.0 & 100 & -156.03748\\
36 & 0.5 & 2.35 & 2.0 & Yes & Flat & 0.0 & 104 & -169.66629\\
37 & 0.1 & 2.7 & 2.0 & Yes & Flat & 0.0 & 8 & -3.8255920\\
38 & 0.2 & 2.7 & 2.0 & Yes & Flat & 0.0 & 38 & -26.668427\\
39 & 0.3 & 2.7 & 2.0 & Yes & Flat & 0.0 & 57 & -48.150604\\
40 & 0.5 & 2.7 & 2.0 & Yes & Flat & 0.0 & 54 & -46.984512\\
41 & 0.1 & 2.35 & 0.25 & Yes & Flat & 0.0 & 164 & -776.80261\\
42 & 0.2 & 2.35 & 0.25 & Yes & Flat & 0.0 & 179 & -1404.6885\\
43 & 0.3 & 2.35 & 0.25 & Yes & Flat & 0.0 & 186 & -2075.5227\\
44 & 0.5 & 2.35 & 0.25 & Yes & Flat & 0.0 & 192 & -3790.4802\\
45 & 0.1 & 2.7 & 0.25 & Yes & Flat & 0.0 & 122 & -270.38574\\
46 & 0.2 & 2.7 & 0.25 & Yes & Flat & 0.0 & 147 & -506.66522\\
47 & 0.3 & 2.7 & 0.25 & Yes & Flat & 0.0 & 162 & -760.54401\\
48 & 0.5 & 2.7 & 0.25 & Yes & Flat & 0.0 & 182 & -1476.1992\\
49 & 0.1 & 2.35 & 1.0 & No & Flat & 0.1 & 74 & -78.037476\\
50 & 0.2 & 2.35 & 1.0 & No & Flat & 0.1 & 108 & -191.24628\\
51 & 0.3 & 2.35 & 1.0 & No & Flat & 0.1 & 132 & -337.98636\\
52 & 0.5 & 2.35 & 1.0 & No & Flat & 0.1 & 154 & -637.43152\\
53 & 0.1 & 2.7 & 1.0 & No & Flat & 0.1 & 22 & -15.239967\\
54 & 0.2 & 2.7 & 1.0 & No & Flat & 0.1 & 58 & -48.793671\\
55 & 0.3 & 2.7 & 1.0 & No & Flat & 0.1 & 78 & -87.307587\\
56 & 0.5 & 2.7 & 1.0 & No & Flat & 0.1 & 94 & -133.84480\\
57 & 0.1 & 2.35 & 2.0 & No & Flat & 0.1 & 30 & -17.080589\\
58 & 0.2 & 2.35 & 2.0 & No & Flat & 0.1 & 68 & -69.020309\\
59 & 0.3 & 2.35 & 2.0 & No & Flat & 0.1 & 91 & -117.72034\\
60 & 0.5 & 2.35 & 2.0 & No & Flat & 0.1 & 86 & -104.91006\\
61 & 0.1 & 2.7 & 2.0 & No & Flat & 0.1 & 7 & -3.7329865\\
62 & 0.2 & 2.7 & 2.0 & No & Flat & 0.1 & 34 & -22.856674\\
63 & 0.3 & 2.7 & 2.0 & No & Flat & 0.1 & 46 & -39.461548\\
64 & 0.5 & 2.7 & 2.0 & No & Flat & 0.1 & 43 & -35.445259\\
65 & 0.1 & 2.35 & 0.25 & No & Flat & 0.1 & 155 & -646.17456\\
66 & 0.2 & 2.35 & 0.25 & No & Flat & 0.1 & 174 & -1132.3506\\
67 & 0.3 & 2.35 & 0.25 & No & Flat & 0.1 & 183 & -1587.7677\\
68 & 0.5 & 2.35 & 0.25 & No & Flat & 0.1 & 189 & -2579.9290\\
69 & 0.1 & 2.7 & 0.25 & No & Flat & 0.1 & 103 & -164.22797\\
70 & 0.2 & 2.7 & 0.25 & No & Flat & 0.1 & 131 & -330.11389\\
71 & 0.3 & 2.7 & 0.25 & No & Flat & 0.1 & 145 & -480.67297\\
72 & 0.5 & 2.7 & 0.25 & No & Flat & 0.1 & 168 & -889.29907\\
73 & 0.1 & 2.35 & 1.0 & Yes & Flat & 0.1 & 88 & -108.42148\\
74 & 0.2 & 2.35 & 1.0 & Yes & Flat & 0.1 & 123 & -275.61371\\
75 & 0.3 & 2.35 & 1.0 & Yes & Flat & 0.1 & 146 & -500.68408\\
76 & 0.5 & 2.35 & 1.0 & Yes & Flat & 0.1 & 171 & -1027.9811\\
77 & 0.1 & 2.7 & 1.0 & Yes & Flat & 0.1 & 27 & -16.603905\\
78 & 0.2 & 2.7 & 1.0 & Yes & Flat & 0.1 & 67 & -68.651489\\
79 & 0.3 & 2.7 & 1.0 & Yes & Flat & 0.1 & 92 & -126.54276\\
80 & 0.5 & 2.7 & 1.0 & Yes & Flat & 0.1 & 117 & -236.25699\\
81 & 0.1 & 2.35 & 2.0 & Yes & Flat & 0.1 & 20 & -13.885048\\
82 & 0.2 & 2.35 & 2.0 & Yes & Flat & 0.1 & 75 & -80.561569\\
83 & 0.3 & 2.35 & 2.0 & Yes & Flat & 0.1 & 95 & -136.36365\\
84 & 0.5 & 2.35 & 2.0 & Yes & Flat & 0.1 & 102 & -161.30212\\
85 & 0.1 & 2.7 & 2.0 & Yes & Flat & 0.1 & 9 & -3.8542175\\
86 & 0.2 & 2.7 & 2.0 & Yes & Flat & 0.1 & 35 & -25.809174\\
87 & 0.3 & 2.7 & 2.0 & Yes & Flat & 0.1 & 53 & -45.016434\\
88 & 0.5 & 2.7 & 2.0 & Yes & Flat & 0.1 & 52 & -44.284752\\
89 & 0.1 & 2.35 & 0.25 & Yes & Flat & 0.1 & 156 & -649.58960\\
90 & 0.2 & 2.35 & 0.25 & Yes & Flat & 0.1 & 177 & -1272.5421\\
91 & 0.3 & 2.35 & 0.25 & Yes & Flat & 0.1 & 185 & -2018.7471\\
92 & 0.5 & 2.35 & 0.25 & Yes & Flat & 0.1 & 191 & -3722.3474\\
93 & 0.1 & 2.7 & 0.25 & Yes & Flat & 0.1 & 107 & -184.78189\\
94 & 0.2 & 2.7 & 0.25 & Yes & Flat & 0.1 & 139 & -405.31018\\
95 & 0.3 & 2.7 & 0.25 & Yes & Flat & 0.1 & 159 & -687.57117\\
96 & 0.5 & 2.7 & 0.25 & Yes & Flat & 0.1 & 180 & -1430.8417\\
193 & 0.1 & 2.35 & 1.0 & No & 50-50 & 0.0 & 109 & -191.83087\\
194 & 0.2 & 2.35 & 1.0 & No & 50-50 & 0.0 & 118 & -243.62000\\
195 & 0.3 & 2.35 & 1.0 & No & 50-50 & 0.0 & 126 & -288.46088\\
196 & 0.5 & 2.35 & 1.0 & No & 50-50 & 0.0 & 138 & -397.25439\\
197 & 0.1 & 2.7 & 1.0 & No & 50-50 & 0.0 & 50 & -42.623398\\
198 & 0.2 & 2.7 & 1.0 & No & 50-50 & 0.0 & 65 & -60.471497\\
199 & 0.3 & 2.7 & 1.0 & No & 50-50 & 0.0 & 70 & -73.344910\\
200 & 0.5 & 2.7 & 1.0 & No & 50-50 & 0.0 & 84 & -97.657059\\
201 & 0.1 & 2.35 & 2.0 & No & 50-50 & 0.0 & 19 & -13.511047\\
202 & 0.2 & 2.35 & 2.0 & No & 50-50 & 0.0 & 45 & -37.678787\\
203 & 0.3 & 2.35 & 2.0 & No & 50-50 & 0.0 & 64 & -60.211609\\
204 & 0.5 & 2.35 & 2.0 & No & 50-50 & 0.0 & 63 & -54.762085\\
205 & 0.1 & 2.7 & 2.0 & No & 50-50 & 0.0 & 1 & 0.0000000\\
206 & 0.2 & 2.7 & 2.0 & No & 50-50 & 0.0 & 14 & -8.9467392\\
207 & 0.3 & 2.7 & 2.0 & No & 50-50 & 0.0 & 25 & -16.300880\\
208 & 0.5 & 2.7 & 2.0 & No & 50-50 & 0.0 & 18 & -12.898163\\
209 & 0.1 & 2.35 & 0.25 & No & 50-50 & 0.0 & 149 & -523.74426\\
210 & 0.2 & 2.35 & 0.25 & No & 50-50 & 0.0 & 160 & -700.39764\\
211 & 0.3 & 2.35 & 0.25 & No & 50-50 & 0.0 & 170 & -943.16479\\
212 & 0.5 & 2.35 & 0.25 & No & 50-50 & 0.0 & 181 & -1438.5382\\
213 & 0.1 & 2.7 & 0.25 & No & 50-50 & 0.0 & 111 & -198.82269\\
214 & 0.2 & 2.7 & 0.25 & No & 50-50 & 0.0 & 121 & -264.05933\\
215 & 0.3 & 2.7 & 0.25 & No & 50-50 & 0.0 & 130 & -328.01355\\
216 & 0.5 & 2.7 & 0.25 & No & 50-50 & 0.0 & 144 & -471.44672\\
217 & 0.1 & 2.35 & 1.0 & Yes & 50-50 & 0.0 & 115 & -217.41257\\
218 & 0.2 & 2.35 & 1.0 & Yes & 50-50 & 0.0 & 124 & -279.08957\\
219 & 0.3 & 2.35 & 1.0 & Yes & 50-50 & 0.0 & 133 & -351.58136\\
220 & 0.5 & 2.35 & 1.0 & Yes & 50-50 & 0.0 & 150 & -558.37683\\
221 & 0.1 & 2.7 & 1.0 & Yes & 50-50 & 0.0 & 60 & -52.107895\\
222 & 0.2 & 2.7 & 1.0 & Yes & 50-50 & 0.0 & 71 & -74.032425\\
223 & 0.3 & 2.7 & 1.0 & Yes & 50-50 & 0.0 & 85 & -99.106094\\
224 & 0.5 & 2.7 & 1.0 & Yes & 50-50 & 0.0 & 96 & -141.66830\\
225 & 0.1 & 2.35 & 2.0 & Yes & 50-50 & 0.0 & 33 & -22.380219\\
226 & 0.2 & 2.35 & 2.0 & Yes & 50-50 & 0.0 & 62 & -53.372421\\
227 & 0.3 & 2.35 & 2.0 & Yes & 50-50 & 0.0 & 73 & -77.036163\\
228 & 0.5 & 2.35 & 2.0 & Yes & 50-50 & 0.0 & 77 & -85.713593\\
229 & 0.1 & 2.7 & 2.0 & Yes & 50-50 & 0.0 & 2 & -0.057250977\\
230 & 0.2 & 2.7 & 2.0 & Yes & 50-50 & 0.0 & 16 & -9.4789734\\
231 & 0.3 & 2.7 & 2.0 & Yes & 50-50 & 0.0 & 31 & -17.890610\\
232 & 0.5 & 2.7 & 2.0 & Yes & 50-50 & 0.0 & 28 & -16.608932\\
233 & 0.1 & 2.35 & 0.25 & Yes & 50-50 & 0.0 & 148 & -510.56641\\
234 & 0.2 & 2.35 & 0.25 & Yes & 50-50 & 0.0 & 167 & -813.52740\\
235 & 0.3 & 2.35 & 0.25 & Yes & 50-50 & 0.0 & 175 & -1156.5618\\
236 & 0.5 & 2.35 & 0.25 & Yes & 50-50 & 0.0 & 188 & -2207.2117\\
237 & 0.1 & 2.7 & 0.25 & Yes & 50-50 & 0.0 & 113 & -206.70996\\
238 & 0.2 & 2.7 & 0.25 & Yes & 50-50 & 0.0 & 128 & -299.82571\\
239 & 0.3 & 2.7 & 0.25 & Yes & 50-50 & 0.0 & 140 & -417.71179\\
240 & 0.5 & 2.7 & 0.25 & Yes & 50-50 & 0.0 & 163 & -771.67303\\
241 & 0.1 & 2.35 & 1.0 & No & 50-50 & 0.1 & 59 & -51.771774\\
242 & 0.2 & 2.35 & 1.0 & No & 50-50 & 0.1 & 83 & -95.560333\\
243 & 0.3 & 2.35 & 1.0 & No & 50-50 & 0.1 & 97 & -147.48036\\
244 & 0.5 & 2.35 & 1.0 & No & 50-50 & 0.1 & 120 & -257.68997\\
245 & 0.1 & 2.7 & 1.0 & No & 50-50 & 0.1 & 4 & -1.6570358\\
246 & 0.2 & 2.7 & 1.0 & No & 50-50 & 0.1 & 23 & -15.514267\\
247 & 0.3 & 2.7 & 1.0 & No & 50-50 & 0.1 & 40 & -29.969849\\
248 & 0.5 & 2.7 & 1.0 & No & 50-50 & 0.1 & 55 & -47.292496\\
249 & 0.1 & 2.35 & 2.0 & No & 50-50 & 0.1 & 10 & -4.1068573\\
250 & 0.2 & 2.35 & 2.0 & No & 50-50 & 0.1 & 36 & -26.424324\\
251 & 0.3 & 2.35 & 2.0 & No & 50-50 & 0.1 & 51 & -43.055374\\
252 & 0.5 & 2.35 & 2.0 & No & 50-50 & 0.1 & 47 & -39.864380\\
253 & 0.1 & 2.7 & 2.0 & No & 50-50 & 0.1 & 5 & -1.7356873\\
254 & 0.2 & 2.7 & 2.0 & No & 50-50 & 0.1 & 13 & -7.6563568\\
255 & 0.3 & 2.7 & 2.0 & No & 50-50 & 0.1 & 21 & -14.323235\\
256 & 0.5 & 2.7 & 2.0 & No & 50-50 & 0.1 & 17 & -11.992485\\
257 & 0.1 & 2.35 & 0.25 & No & 50-50 & 0.1 & 135 & -366.92471\\
258 & 0.2 & 2.35 & 0.25 & No & 50-50 & 0.1 & 153 & -583.01910\\
259 & 0.3 & 2.35 & 0.25 & No & 50-50 & 0.1 & 166 & -804.36145\\
260 & 0.5 & 2.35 & 0.25 & No & 50-50 & 0.1 & 178 & -1347.9231\\
261 & 0.1 & 2.7 & 0.25 & No & 50-50 & 0.1 & 81 & -92.800415\\
262 & 0.2 & 2.7 & 0.25 & No & 50-50 & 0.1 & 98 & -153.29453\\
263 & 0.3 & 2.7 & 0.25 & No & 50-50 & 0.1 & 114 & -212.33896\\
264 & 0.5 & 2.7 & 0.25 & No & 50-50 & 0.1 & 137 & -392.63898\\
265 & 0.1 & 2.35 & 1.0 & Yes & 50-50 & 0.1 & 79 & -88.956650\\
266 & 0.2 & 2.35 & 1.0 & Yes & 50-50 & 0.1 & 99 & -153.70737\\
267 & 0.3 & 2.35 & 1.0 & Yes & 50-50 & 0.1 & 116 & -235.27686\\
268 & 0.5 & 2.35 & 1.0 & Yes & 50-50 & 0.1 & 143 & -447.87061\\
269 & 0.1 & 2.7 & 1.0 & Yes & 50-50 & 0.1 & 12 & -5.6114807\\
270 & 0.2 & 2.7 & 1.0 & Yes & 50-50 & 0.1 & 39 & -27.452682\\
271 & 0.3 & 2.7 & 1.0 & Yes & 50-50 & 0.1 & 56 & -47.319824\\
272 & 0.5 & 2.7 & 1.0 & Yes & 50-50 & 0.1 & 76 & -83.936432\\
273 & 0.1 & 2.35 & 2.0 & Yes & 50-50 & 0.1 & 6 & -2.3401947\\
274 & 0.2 & 2.35 & 2.0 & Yes & 50-50 & 0.1 & 41 & -31.392036\\
275 & 0.3 & 2.35 & 2.0 & Yes & 50-50 & 0.1 & 61 & -52.974701\\
276 & 0.5 & 2.35 & 2.0 & Yes & 50-50 & 0.1 & 66 & -63.390015\\
277 & 0.1 & 2.7 & 2.0 & Yes & 50-50 & 0.1 & 3 & -0.53945923\\
278 & 0.2 & 2.7 & 2.0 & Yes & 50-50 & 0.1 & 15 & -9.2003784\\
279 & 0.3 & 2.7 & 2.0 & Yes & 50-50 & 0.1 & 24 & -16.015152\\
280 & 0.5 & 2.7 & 2.0 & Yes & 50-50 & 0.1 & 26 & -16.512665\\
281 & 0.1 & 2.35 & 0.25 & Yes & 50-50 & 0.1 & 134 & -364.58423\\
282 & 0.2 & 2.35 & 0.25 & Yes & 50-50 & 0.1 & 157 & -666.35565\\
283 & 0.3 & 2.35 & 0.25 & Yes & 50-50 & 0.1 & 172 & -1048.5062\\
284 & 0.5 & 2.35 & 0.25 & Yes & 50-50 & 0.1 & 187 & -2083.5793\\
285 & 0.1 & 2.7 & 0.25 & Yes & 50-50 & 0.1 & 87 & -105.98897\\
286 & 0.2 & 2.7 & 0.25 & Yes & 50-50 & 0.1 & 110 & -193.32385\\
287 & 0.3 & 2.7 & 0.25 & Yes & 50-50 & 0.1 & 129 & -315.21362\\
288 & 0.5 & 2.7 & 0.25 & Yes & 50-50 & 0.1 & 158 & -687.04688

\enddata

\tablenotetext{a}{CE efficiency parameter}
\tablenotetext{b}{Stellar wind strength parameter}
\tablenotetext{c}{0: CE from Hertzsprung gap donor allowed, 1: not allowed}
\tablenotetext{d}{binary mass ratio distribution.}
\tablenotetext{e}{SN kicks given to direct collapse black holes.  0.0 = no SN kick given, 0.1 = small SN kick given}
\tablenotetext{f}{The rank of the model based on the likelihood value}
\tablenotetext{g}{Log of the ratio of the likelihood of the given model to that of the highest likelihood model}
\end{deluxetable}

\end{document}